\documentclass[12pt]{article}
\usepackage{amsfonts}
\usepackage[utf8]{inputenc}
\usepackage{bm}
\usepackage{amsmath}
\usepackage{amssymb}
\usepackage{epsfig}
\usepackage{enumitem}
\usepackage{cite}
\usepackage[bookmarks=true,colorlinks=true,linkcolor=black,citecolor=orange,urlcolor=orange,bookmarksnumbered]{hyperref}
\usepackage{cancel}

\usepackage[dvipsnames]{xcolor}

\setlist[enumerate]{%
wide =0.5\parindent,
listparindent=0pt%
}

\newcommand{\bmat}{\left(\begin{array}}
\newcommand{\emat}{\end{array}\right)}

\def\gtrsim{\mathrel{\raise.3ex\hbox{$>$\kern-.75em\lower1ex\hbox{$\sim$}}}}

\def\-{\hphantom{-}}

\def\s2{\frac{1}{\sqrt2}}

\def\mg{m_{3/2}}
\def\mg2{m^2_{3/2}}

\def\Dsl{\,\raise.15ex\hbox{/}\mkern-13.5mu D} 

\def\be{\begin{equation}}
\def\ee{\end{equation}}
\def\bea{\begin{eqnarray}}
\def\eea{\end{eqnarray}}

\newcommand{\nn}{\nonumber}

\begin{document}


\pagestyle{plain}

\makeatletter
\@addtoreset{equation}{section}
\makeatother
\renewcommand{\theequation}{\thesection.\arabic{equation}}
\pagestyle{empty}
\begin{center}
\ \

\vskip .5cm

\LARGE{\bf A Non-Relativistic Limit for Heterotic Supergravity and its Gauge Lagrangian} \\[10mm]

\vskip 0.3cm

\large{Eric Lescano
 \\[6mm]}

{\small Institute for Theoretical Physics (IFT), University of Wroclaw, \\
pl. Maxa Borna 9, 50-204 Wroclaw,
Poland\\ [.01 cm]}

{\small \verb"{eric.lescano}@uwr.edu.pl"}\\[1cm]

\small{\bf Abstract} \\[0.5cm]

 \end{center}

Motivated by the recent construction of non-relativistic (NR) heterotic Double Field Theory (HDFT), we analyze the $D=10$ bosonic sector of heterotic supergravity under a consistent NR limit. We show that the resulting theory admits a finite Lagrangian due to non-trivial cancellations of divergent contributions arising from the Chern–Simons terms in the curvature of the $\hat B$-field and from the Yang–Mills sector. This mechanism parallels the well-known cancellation between the Ricci scalar $\hat R$ and the $-\frac{1}{12}\hat H^2$ term in bosonic supergravity under the same limit. Extending previous analyses, we incorporate the full bosonic gauge sector emerging from the HDFT construction and derive the complete finite bosonic heterotic Lagrangian in manifestly gauge-covariant form, written in terms of gauge-covariant curvatures and derivatives. An interesting feature of the NR expansion is that the gauged Green–Schwarz transformation of the two-form trivializes (it can be eliminated by imposing field redefinitions), while terms equivalent to Chern–Simons contributions naturally re-emerge in the effective theory.   


\setcounter{page}{1}
\pagestyle{plain}
\renewcommand{\thefootnote}{\arabic{footnote}}
\setcounter{footnote}{0}
\newpage
\tableofcontents 
\newpage
\section{Introduction}

In the last decade, there has been significant progress made in the study of different formulations of non-relativistic string theory \cite{NR1}-\cite{NR3} and its supergravity limit \cite{NRST1}-\cite{NRST13} \footnote{See \cite{Review}-\cite{Review2} for reviews, and also \cite{Review3} for a complementary introduction to this topic.}. The low-energy limit of bosonic string theory is given by the universal NS-NS supergravity, whose fundamental fields $\hat g$, $\hat B$, $\hat \phi$ admit a non-relativistic expansion of the form \cite{NSNS},
\bea
    \hat{g}_{\mu \nu} & = & c^2 {\tau_\mu}^a {\tau_\nu}^b \eta_{ab} + h_{\mu \nu} \, , \label{metricexpansion} \\
    \hat{g}^{\mu \nu} & = & \frac{1}{c^2} {\tau^\mu}_a {\tau^\nu}_b \eta^{ab} + h^{\mu \nu} \, , \label{inversemetricexpansion} \\
    \hat{B}_{\mu \nu} & = & - c^2 {\tau_\mu}^a {\tau_\nu}^{b}  \epsilon_{ab} + b_{\mu \nu} \label{Bexpansion}\, ,  \\
\hat{\phi} & = & \ln(c) + \varphi \, ,
\eea
where $\mu,\nu=1,\dots,10$ are space-time indices and we have split the flat index $\hat a=(a,a')$ where $a,b, ...=0,1$ are the longitudinal directions and $a^\prime,b^\prime,...=2,\dots,9$. The quantities in the previous expansion obey the following Newton-Cartan relations,
\begin{align}
  \tau_{\mu}{}^{a} e^{\mu}{}_{a'} & = \tau^{\mu}{}_{a} e_{\mu}{}^{a'} = 0 \, , \qquad  e_{\mu}{}^{a'} e^{\mu}{}_{b'} = \delta^{a'}_{b'} \, , \\
  \tau_{\mu}{}^{a} \tau^{\mu b} & = \eta^{a b} \, , \qquad \tau_{\mu}{}^{a} \tau^{\nu}{}_{a} + e_{\mu}{}^{a'} e^{\nu}{}_{a'} = \delta_{\mu}^{\nu} \, ,
\label{NCrelations}
\end{align} 
where $h_{\mu \nu}=e_{\mu}{}^{a'} \eta_{a' b'} e_{\nu}{}^{b'}$ and $h^{\mu \nu}=e^{\mu}{}_{a'} \eta^{a' b'} e^{\nu}{}_{b'}$. While, a priori, all these expansions produce divergent tensorial quantities upon taking the NR limit $c \rightarrow \infty$ (i.e. Riemann tensor and its traces, H strength, etc.), the NS-NS supergravity Lagrangian,
\bea
S = \int d^{10}x \sqrt{-\hat g} e^{-2 \hat \phi} \Big( \hat R + 4 \partial_{\mu} \hat \phi \partial^{\mu} \hat \phi - \frac{1}{12} \hat H_{\mu \nu \rho} \hat H^{\mu \nu \rho}\Big)  \, ,
\label{Sbos0}
\eea
with 
\bea
\hat H_{\mu \nu \rho} =  3 \partial_{[\mu} \hat B_{\nu \rho]} \, ,
\eea
remains finite due to a curious cancellation between the divergent part of the Ricci scalar and the divergent part of the $\hat H$ squared term \cite{NSNS}. 

In the case of the bosonic sector of heterotic supergravity \footnote{The present work is restricted to the bosonic sector of heterotic supergravity. For readers interested in the fermionic sector, they were included in \cite{NRST11}, where the authors analyzed the divergences of the supersymmetric transformations.}, we expect a similar cancellation when including the gauge field $\hat A_{\mu}^{i}$, $i=1,\dots,N$ and $N=496$ which is the dimension of the heterotic gauge group. Given a particular expansion for the gauge field, the inclusion of Chern-Simons terms in the curvature of the $\hat B$-field should produce divergences that must be compensated by the Yang-Mills Lagrangian. A first expansion for the $A$-field was constructed in \cite{BandR}, where the authors showed the cancellation of divergences due to the new gauge contributions. Particularly, the authors expanded the non-Abelian gauge field according to
\bea
A_{\nu}{}^{i} = c^2  \tau_{\nu}{}^{-} v_{-}{}^{i} +  \tau_{\nu}{}^{+} v_{+}{}^{i} +  e_{\nu}{}^{a'} v_{a'}{}^{i} \, .
\label{BRVexpansion}
\eea
In this formalism, the c-expansion of the vielbein includes gauge field contributions given by  
\bea
\hat E_{\mu}{}^{-} & = & c  \tau_{\mu}{}^{-}, \\ \hat E_{\mu}{}^{+} & = & - c^3 \frac{v_{-}{}^{i} v_{- i}}{2}  \tau_{\mu}{}^{-} + c  \tau_{\mu}{}^{+} \, , \\ \hat E_{\mu}{}^{a'} & = &  e_{\mu}{}^{a'} \, ,
\eea
while the expansion for the B-field is also deformed 
\bea
\hat B_{\mu \nu} = c^2  \tau_{\mu}{}^a { \tau_\nu}^{b}  \epsilon_{ab} + 2 c^2  \tau_{[\mu}{}^{-}  e_{\nu]}{}^{a'} v_{-}{}^{i} v_{a' i} +  b_{\mu \nu} \, .
\eea

The gauge transformations of the gauge degrees of freedom are given by
\bea
\delta_{\lambda} v_{+}{}^{i} & = & \partial_{+} \lambda^{i} + \sqrt{2} f_{jk}{}^{i} \lambda^{j} v_{+}{}^{k} \, , \\
\delta_{\lambda} v_{-}{}^{i} & = & \sqrt{2} f_{jk}{}^{i} \lambda^{j} v_{-}{}^{k} \, , \\
\delta_{\lambda} v_{a'}{}^{i} & = & \partial_{a} \lambda^{i} + \sqrt{2} f_{jk}{}^{i} \lambda^{j} v_{a'}{}^{k} \, , 
\eea
where $\lambda$ is an arbitrary gauge parameter. From the previous equation, we observe that $v_{+}{}^{i}$ and $v_{a}{}^{i}$ play the role of connections, while $v_{-}{}^{i}$ transforms like a gauge vector. In this prescription $ \tau_{-}{}^{a}$ is gauge invariant but $ \tau_{+}{}^{a}$ and $b_{\mu \nu}$ transform in a non-covariant way under gauge transformations \cite{BandR},
\bea
\label{noncov1}
\delta_{\lambda}  \tau_{\mu}{}^{+} & = & \frac{v_{-}{}^{i} \partial_{-} \lambda^{i}}{1+v_{+}{}^{i} v_{-i}}  \tau_{\mu}{}^{-} \, , \\
\delta\,  b_{\mu \nu}&=&2\tau_{[\mu}{}^{+}e_{\nu]}{}^{a}\Big(v_{+I}\partial_{a}\lambda^{I}-v_{ai}\partial_{+}\lambda^{i}\Big)-2e_{[\mu}{}^{a}e_{\nu]}{}^{b}v_{a}^{I}\partial_{b}\lambda_{i}+\nn \\
&& +2\bigg[\tau_{[\mu}{}^{-}\tau_{\nu]}{}^{+}\bigg(-2v_{+i}+\frac{v_{+}^{2}}{1+v_{+-}}v_{-i}\bigg)+\tau_{[\mu}{}^{-}e_{\nu]}{}^{a}\bigg(-2v_{ai}+\frac{v_{a+}}{1+v_{+-}}v_{-i}\bigg)\bigg]\partial_{-}\lambda^{i} \, . \nn
 \\ \label{noncov2}
\eea
Due to the complexity of the previous transformations, it is natural to wonder if there could exist a set of field redefinitions that could help us to partially simplify them. Moreover, it was recently proven that the non-relativistic Lorentz Green-Schwarz transformation of the B-field can be trivialized (independently of the expansion of the gauge sector) by imposing field redefinitions \cite{Trivialization}. Since the Lorentz and Lorentz Green-Schwarz mechanism are related to guaranteeing the cancellation of anomalies, it would be desirable to have a similar result for the gauge sector. Then, it is natural to inspect whether the transformations (\ref{noncov1})-(\ref{noncov2}) can be trivialized or, alternatively, if there exists a different expansion of the non-Abelian gauge field which might have a trivial Green-Schwarz gauge mechanism. Particularly it would be desirable to have a formulation where $\delta  \tau_{\mu}{}^{+}=0$, so that the source of non-covariance could be redefined as just an A-dependent field redefinition (we elaborate on this in section 4.3). Alternatively to (\ref{BRVexpansion}), another possible expansion to address the non-relativistic limit of the bosonic sector of the heterotic supergravity was given in \cite{EandD}, using a manifest T-duality scenario. This formulation provides a finite heterotic Double Field Theory Lagrangian, guaranteed by the construction of a finite generalized metric and generalized dilaton.  Since T-duality is an exact symmetry of string theory, the NR limit of the heterotic supergravity admits a rewriting in terms of $O(D,D+N)$ multiplets \cite{EandD} within the framework of Double Field Theory (DFT)\cite{DFT1}-\cite{DFT4}, when the fundamental degrees of freedom are parametrized using a non-Riemannian ansatz \cite{NRDFT1}-\cite{NRDFT4}. 

In the ungauged case, the fundamental fields of DFT are given by a generalized metric, ${\cal H}(\hat g, \hat b)$ , and a generalized dilaton $\hat d$. The non-relativistic expansion of these fields is given by
\bea
\hat {\cal H}_{M N} & = &  \hat{\cal H}^{(0)}_{M N} + \frac{1}{c^2} \hat {\cal H}^{(-2)}_{M N} \, \\
e^{-2 \hat{d}} & = & e^{-2\hat{\Phi}} \sqrt{-\hat g} = e^{-2 \phi} \sqrt{f(\tau,h)} = e^{-2 d} \, ,
\eea
where $f(\tau,h) = - \frac{\hat{g}}{c^4}$ and the finite generalized metric has the form
\begin{align}
    {\cal H}^{(0)}_{M N} &=  
\left(\begin{matrix} h^{\mu \nu} &  \epsilon_{a b} \tau_{\nu}{}^{b} \tau^{\mu a}  - b_{\rho \nu} h^{\rho \mu} \\ 
\epsilon_{a b} \tau_{\mu}{}^{b} \tau^{\nu a}  - b_{\rho \mu} h^{\rho \nu} &  h_{\mu \nu} + b_{\rho \mu} h^{\rho \sigma} b_{\sigma \nu} - 2 \epsilon_{c d}  \tau_{(\mu|}{}^{d} b_{\sigma |\nu)} \tau^{\sigma c} \, \end{matrix} \right) \, ,
\end{align}
where $M,N...$ are indices in the fundamental representation of $O(D,D)$.

While in the supergravity limit the metric $\hat g_{\mu \nu}$ contains a $c^2$ contribution, in (bosonic) DFT the generalized metric eliminates this term due to an interplay between $\hat g_{\mu \nu}$ and the $\hat B$-field. Therefore the convergence of the DFT Lagrangian ${\cal R}(\hat{\mathcal H}, e^{-2\hat d})$ and the DFT equations of motion is automatically guaranteed.

The extension from bosonic DFT to heterotic DFT can be easily done by extending the duality group from $O(D,D)$ to $O(D,D+N)$, where $N$ is the dimension of the heterotic gauge group. The components of the generalized metric in this case contain a metric $\hat g_{\mu \nu}$, a $\hat B$-field and a gauge field $\hat A_{\mu}{}^{i}$, which transforms as a connection under the gauge group. The expansion (\ref{BRVexpansion}) proposed by Bergshoeff and Romano (BR), together with the deformation of the vielbein and the B-field leads to a divergent generalized metric,
\bea
&{}&  \hat{\cal H}_{\cal M N}(\hat{g},\hat{B},\hat{A})|_{BR} \label{eq:BRExpansionGeneralisedMetric}\\
&=& c^4 \ {\cal H}^{(4)}_{\cal M N} (\tau,h,b,\alpha,a)  + c^2 \ {\cal H}^{(2)}_{\cal M N}(\tau,h,b,\alpha,a)  + {\cal H}^{(0)}_{\cal M N}(\tau,h,b,\alpha,a) + \mathcal{O}\left(\frac{1}{c^2}\right) \, . \nonumber
\eea

While in principle this is not a major problem, there exist other expansions which lead to a finite generalized metric and simple transformations rules (all the conditions for obtaining a finite generalized metric were studied in \cite{EandD}). One possibility is to expand the field $\hat A_{\mu}{}^{i}$ following the proposal 
 \bea
 \hat{A}_{\mu}^{i} &= c \ {\tau_\mu}^{-} \alpha_{-}^i + \frac{1}{c} a_\mu^i \, .
\label{Aexpansion}
\eea
 The expansion (\ref{Aexpansion}), proposed by Lescano and Osten (LO) was constructed in such a way that the generalized metric of heterotic DFT, which is an element of $O(D,D+N)$, remains convergent after taking the NR limit in the double geometry,
 \begin{align}
   &{} \qquad \hat{\cal H}_{\cal M N}(\hat{g},\hat{B},\hat{A})|_{LO} \label{eq:HeteroticGeneralisedMetricFinite}  \\
   &= {\mathcal{H}}^{(0)}_{\cal MN} (\tau,h,b,\alpha,a) + \frac{1}{c} {\mathcal{H}}^{(-1)}_{\cal MN} (\tau,h,b,\alpha,a) + \frac{1}{c^2} {\mathcal{H}}^{(-2)}_{\cal MN} (\tau,h,b,\alpha,a) + \mathcal{O}\left( \frac{1}{c^3} \right) \nonumber
\end{align}
In this way, one can use manifest T-duality to obtain a simpler non-relativistic theory at the supergravity level. Moreover, DFT ensures the finiteness of the non-Abelian gauge transformations since they are encoded in the generalized Lie derivative of the generalized metric, as shown in \cite{EandD}. 
 
 Demanding this alternative expansion for the A-field allows one to have a finite action and equations of motion at the DFT level, by construction, upon considering the NR limit, since both the Lagrangian and the equations of motion depend on the generalized metric and dilaton (metric formulation of DFT). Therefore in this work we focus on studying the non-relativistic limit of the bosonic sector of the heterotic supergravity when one imposes (\ref{Aexpansion}) instead of (\ref{BRVexpansion}). The immediate advantage, besides that the theory produces a finite generalized metric, is that the $\tau^{\mu}_{+}$ does not transform under gauge transformations and the Green-Schwarz mechanism is trivial (main result of the paper). Also, we will explicitly obtain the finite Lagrangian after imposing $c\rightarrow \infty$, obtained from
\bea
S_{\rm{het}} = \int d^{10}x \sqrt{-\hat g} e^{-2 \hat \phi} (\hat R + 4 \partial_{\mu} \hat \phi \partial^{\mu} \hat \phi - \frac{1}{12} \hat{\bar{H}}_{\mu \nu \rho} \hat{\bar{H}}^{\mu \nu \rho} - \frac14 \hat F_{\mu \nu}{}^{i} \hat F^{\mu \nu}{}_{i}) \, ,
\label{Shet0}
\eea
where
\be
\bar H_{\mu\nu\rho}=3\left(\partial_{[\mu}\hat B_{\nu\rho]}- \hat C_{\mu\nu\rho}^{(g)}\right)\, , \label{barH}
\ee 
\be
\hat C_{\mu\nu\rho}^{(g)}= \hat A^i_{[\mu}\partial_\nu \hat A_{\rho]i}-\frac13 \hat f_{ijk} \hat A_\mu^i \hat A_\nu^j \hat A_\rho^k \, ,
\ee
\bea
\hat F_{\mu \nu}{}^{i} = 2 \partial_{[\mu} \hat A_{\nu]}{}^{i} - \hat f^{i}{}_{jk} \hat A_{\mu}{}^{j} \hat A_{\nu}{}^{k} \, ,
\eea
In order to present our results in a compact way, we will write the finite Lagrangian in its manifest form with respect to the non-Abelian gauge transformations.  

\subsection{Organization of the paper}

This paper is organized as follows: in the next section we review the cancellation of divergences in the ungauged sector of the heterotic supergravity (without fermions) and then we show the cancellation of divergences in the gauged sector by including (\ref{Aexpansion}). In section \ref{Transformations} we study the gauge transformation rules of the fundamental fields of the theory upon considering the NR limit, and we construct a suitable gauge covariant derivative and gauge curvatures for the gauge fields. In section \ref{Lagrangian} we compute the finite Lagrangian of the theory in its gauge covariant form and, finally, in section \ref{Conclusions}, we present a discussion and an outlook.

\section{Cancellation of divergences}
\label{Cancellation}

When one expands the action (\ref{Sbos0}), both the $\hat R$ and $-\frac{1}{12} \hat H^2$ terms produce $c^2$ contributions. The same happens when we analyze (\ref{Shet0}), but this time the gauge compensation of $- \frac{1}{12} \bar H^2$ is given also by the $-\frac{1}{4} \hat F^2$ term. We will analyze both scenarios separately. 
\subsection{Ungauged case (review)}
Let us review the construction given in \cite{NSNS}, with a focus on the cancellation of divergences in the bosonic supergravity Lagrangian. To do so, we define
\bea
\hat R^{\rho}{}_{\sigma \mu \nu} = - 2 \partial_{[\mu} \Gamma^{\rho}_{\nu] \sigma} - 2 \Gamma^{\rho}_{[\mu| \lambda} \Gamma^{\lambda}_{|\nu] \sigma}
\eea
and $\hat R= \hat R^{\rho}{}_{\mu \nu \sigma} \delta_{\rho}^{\nu} \hat g^{\mu \sigma}$. The divergent part of the Ricci scalar, after imposing the Newton-Cartan relations (\ref{NCrelations}), is given by
\bea
\hat R^{(2)} = - \frac{c^2}{2} \eta_{a b} \partial_{\mu}{\tau_{\nu}{}^{a}}  \partial_{\rho}{\tau_{\sigma}{}^{b}} (h^{\mu \sigma} h^{\nu \rho} - h^{\mu \rho} h^{\nu \sigma}) \, .
\eea
So far we have just considered the expansion of the metric tensor (\ref{metricexpansion}) and its inverse (\ref{inversemetricexpansion}). Therefore, for backgrounds where $h^{\mu \nu}$ is constant, the divergence in the Ricci scalar becomes trivial and it is possible to define a non-relativistic supergravity without including a non-relativistic expansion in the $\hat B$-field. 

For a more general scenario we impose (\ref{Bexpansion}) and then the divergent part of $-\frac{1}{12} \hat H^2$ is given by
\bea
- \frac{1}{12} (\hat H_{\mu \nu \rho} \hat H^{\mu \nu \rho})^{(2)}  = - \frac{c^2}{2} \epsilon_{a b} \epsilon_{c d} \eta^{a c} \partial_{\mu}\tau_{\nu}{}^{b} \partial_{\rho}\tau_{\sigma}{}^{d} (h^{\mu \rho} h^{\nu \sigma} - h^{\mu \sigma} h^{\nu \rho}) \, .
\eea
Using the relation $\epsilon_{a b} \epsilon_{c d} \eta^{a c} = \eta_{b d}$ one can cancel the divergences coming from the Ricci scalar term.

\subsection{Gauged case}
The incorporation of the gauge field $\hat A_{\mu}^{i}$ brings new divergences to the supergravity Lagrangian. On the one hand, the Yang-Mills Lagrangian,
\bea
L_{\textrm{YM}} = -\frac{1}{4} \hat F_{\mu \nu}^{i} \hat F^{\mu \nu}_{i} 
\eea
upon considering the non-relativistic expansions (\ref{Aexpansion}) and (\ref{inversemetricexpansion}), contains a divergence of the form (see appendix \ref{ap+} for intermediate steps)
\bea
L_{\textrm{YM}}^{(2)} = - \frac{c^2}{2} \alpha_{-}^{i} \alpha_{- i} \partial_{\mu}{\tau_{\nu}^{-}} \partial_{\rho}{\tau_{\sigma}^{-}} (h^{\mu \rho} h^{\nu \sigma} - h^{\mu \sigma} h^{\nu \rho}) \, .
\label{YMdiv}
\eea
These terms indicate that the $A$-expansion in (\ref{Aexpansion}) does not lead to a finite Yang-Mills theory. However, the inclusion of Chern-Simons terms in the heterotic Lagrangian (\ref{Shet0}) compensates the divergence (\ref{YMdiv}). Indeed, one can compute (we have used CADABRA \cite{Cadabra} for performing this test, but the main details of the cancellation are given in the Appendix \ref{ap+}), 
\bea
- \frac{1}{12} (\bar H_{\mu \nu \rho} \bar H^{\mu \nu \rho})^{(2)} = - R^{(2)} +\frac{1}{4} (\hat F_{\mu \nu}^{i} \hat F^{\mu \nu}_{i})^{(2)} \, .    
\label{barH2cancel}
\eea
At this point, we have proved that the heterotic supergravity Lagrangian is finite, but trying to write it in a non-covariant form leads to a huge number of terms. For this reason, our strategy will be to write it in  a gauge covariant form, making use of the gauge invariance of the action.

In the next section, we will study the form of the gauge transformation rules and we will construct a suitable gauge covariant derivative. After that, we will construct new curvatures for all the fundamental fields transforming in a non-covariant way under gauge transformations. 

\section{Transformations and curvatures}
In this section we will focus on the non-Abelian gauge transformations, in order to be able to write a gauge covariant action using tensors under this symmetry. We will start by describing the Levi-Civita connection, then we will construct the gauge covariant derivative and finally we will include gauge transformations with a gauging in the covariant derivative. Remarkably, we will describe the Green-Schwarz mechanism for the b-field and its trivialization, considering field redefinitions.
\label{Transformations}

\subsection{Levi-Civita connection}
The Levi-Civita connection can be written as
\bea
\Gamma_{\mu \nu}^{\rho} = c^2 \Gamma_{\mu \nu}^{(2)\rho} + \Gamma_{\mu \nu}^{(0)\rho} + \frac{1}{c^2} \Gamma_{\mu \nu}^{(-2)\rho} \, , 
\eea
where the explicit form of the previous terms is given by
\bea
\Gamma_{\mu \nu}^{(2)\rho} & = & h^{\rho \sigma} (\partial_{\mu} \tau_{\sigma}{}^{a} - \partial_{\sigma} \tau_{(\mu}{}^{a}) \tau_{\nu)}{}^{b} \eta_{a b} \, , \\
\Gamma_{\mu \nu}^{(0)\rho} & = & \frac12 h^{\rho \sigma} (2 \partial_{(\mu} h_{\nu) \sigma} - \partial_{\sigma} h_{\mu \nu}) + \frac12 \tau^{\rho a} \tau^{\sigma}{}_{a} (2 \partial_{(\mu} (\tau_{\nu)}{}^{b} \tau_{\sigma b}) - \partial_{\sigma}(\tau_{\mu}{}^{b} \tau_{\nu b})) \, , \\
\Gamma_{\mu \nu}^{(-2)\rho} & = & - \partial_{( \mu} \tau^{\sigma a} h_{\nu ) \sigma} \tau^{\rho}{}_{a} - \tau^{\rho a} \tau^{\sigma}{}_{a} \partial_{\sigma} h_{\mu \nu} \, . 
\eea
Inspecting the transformation rules under infinitesimal diffeomorphisms given by the Lie derivative $L_{\xi}$, i.e.,
\bea
\delta_{\xi} h_{\mu \nu} & = & L_{\xi} h_{\mu \nu} = \xi^{\rho} \partial_{\rho} h_{\mu \nu} + 2 \partial_{(\mu} \xi^{\rho} h_{\nu) \rho} \, , \\ 
\delta_{\xi} h^{\mu \nu} & = & L_{\xi} h^{\mu \nu} = \xi^{\rho} \partial_{\rho} h^{\mu \nu} - 2 \partial_{\rho} \xi^{(\mu} h^{\nu)\rho}  \, , \\  
\delta_{\xi} \tau_{\mu}{}^{a} & = & L_{\xi} \tau_{\mu}{}^{a} = \xi^{\rho} \partial_{\rho} \tau_{\mu}{}^{a} +  \partial_{\mu} \xi^{\rho} \tau_{\rho}{}^{a}  \, , \\
\delta_{\xi} \tau^{\mu}{}_{a} & = & L_{\xi} \tau^{\mu}{}_{a} = \xi^{\rho} \partial_{\rho} \tau^{\mu}{}_{a} - \partial_{\rho} \xi^{\mu} \tau^{\rho}{}_{a} \, ,
\eea
where $\xi^{\mu}$ is an arbitrary parameter. It is straightforward to check that $\Gamma_{\mu \nu}^{(2)\rho}$ and $\Gamma_{\mu \nu}^{(-2)\rho}$ transform covariantly under infinitesimal diffeomorphisms and they can be interpreted as covariant shifts of the non-covariant part of the non-relativistic connection, given by $\Gamma_{\mu \nu}^{(0)\rho}$. 
\subsection{Covariant derivative}
The covariant derivative of $\tau_{\nu}{}^{a}$ can be easily computed, 
\bea
\label{covT}
\nabla_{\mu} \tau_{\nu}{}^{a} & = & \partial_{\mu} \tau_{\nu}{}^{a} - \Gamma_{\mu \nu}^{\rho} \tau_{\rho}{}^{a} \\ & = & \partial_{\mu} \tau_{\nu}{}^{a} + T_{\mu \nu}^{a} \, , \nn 
\eea
where $T_{\mu \nu}{}^{a} = T_{\nu \mu}{}^{a}$ is given by the following quantity,
\bea
T_{\mu \nu}{}^{a} & = & + \partial_{(\mu}{\tau_{\nu)}{}^{a}}  + \eta_{b c} \eta^{a d} \nabla_{(\mu|}{\tau_{\rho}{}^{b}} \tau_{|\nu)}{}^{c} \tau^{\rho}{}_{d} - \eta_{b c} \eta^{a d} \nabla_{\rho}{\tau_{(\mu|}{}^{b}} \tau_{|\nu)}{}^{c} \tau^{\rho}{}_{d} \nn \\ && - \frac{1}{c^2} \partial_{(\mu}{\tau_{\nu)}{}^{a}} - \frac{1}{c^2} \eta_{b c} \eta^{a d} \nabla_{(\mu}{\tau_{\rho}{}^{b}} \tau_{\nu)}{}^{c} \tau^{\rho}{}_{d} + \frac{1}{c^2} \eta_{b c} \eta^{a d} \nabla_{\rho}{\tau_{(\mu}{}^{b}} \tau_{\nu)}{}^{c} \tau^{\rho}{}_{d} \, . 
\eea
An interesting observation of $\nabla_{\mu} \tau_{\nu}{}^{a}$ is that this quantity does not diverge, but contains $\frac{1}{c^2}$ terms that, in general, one needs to track in order to correctly write a finite Lagrangian. Moreover, the antisymmetric part of $\partial_{\mu} \tau_{\nu}^{a}$ satisfies 
\bea
\nabla_{[\mu} \tau_{\nu]}^{a} & = & \partial_{[\mu} \tau_{\nu]}{}^{a} \, .
\label{antis}
\eea
Therefore, the antisymmetric part of $\partial_{\mu} \tau_{\nu}^{a}$ is covariant under diffeomorphims, giving rise to the well-known intrinsic torsion of the theory \cite{NSNS}. In order to write the two-derivative ungauged action in a covariant way we will just need $R^{(0)}$ given in (\ref{Scalar}) (which we prove in the appendix that it is a covariant quantity), and the relation (\ref{antis}) to replace ordinary derivatives by covariant derivatives in some antisymmetric expressions in the terms coming from the $\bar H^2$ contribution.

\subsection{Gauge transformations}
Following \cite{EandD}, we start by imposing the following rescaling on the gauge parameter and the structure constants,
\bea
\hat \lambda^i & = & \frac{1}{c} \lambda^i \\
\hat f_{ijk} & = & c f_{ijk} \, . 
\label{scaling}
\eea
It is important to stress that $\hat f_{ijk}$ are the structure constants which gauge the generalized Lie derivative of NR heterotic DFT \cite{EandD}, and therefore they are not in principle identified with any gauge group. The scaling (\ref{scaling}) should therefore be interpreted as a contraction of the gauge algebra adapted to the non-relativistic limit, rather than as a rescaling of fixed physical structure constants. In other words, the gauge algebra of NR heterotic DFT is singular in the double geometry \footnote{A similar behavior involving a singular gauge group in relativistic limit of Heterotic Double Field Theory happens when $\alpha'$-corrections are introduced by the a generalized Bergshoeff-de Roo identification. For more details see \cite{gbdr}.} when $c\rightarrow\infty$, and produces a regular heterotic algebra after the rescaling, meaning that $f_{ijk}$ are $E8\times E8$ or $SO(32)$ structure constants. This is a pure non-relativistic effect, for relativistic heterotic DFT one can claim  that the structure constants at the level of the double geometry and their parametrization matches. 

On the other hand, one can think that the rescaling of the structure constants might include divergences in the gauge Lagrangian. However, this rescaling is safe because higher order $\hat f_{ijk}$ contributions are $\propto \frac{1}{c}$. Then, the rescaling only introduces cubic contributions in gauge fields in the finite Lagrangian, which we will present in the next section. 

We find the following rules when taking the limit $c\rightarrow \infty$, 
\bea
\delta_{\lambda} a_\mu^i & = & \partial_\mu \lambda^i + f^i{}_{jk} \lambda^j a_\mu^k\, , \label{gaugetransa} \\
\delta_{\lambda} \alpha_{-}^i & = & f^i{}_{jk} \lambda^j \alpha_{-}^k\, , \label{gaugetransalpha} \\ 
\delta_{ \lambda} b_{\mu\nu} & = & -(\partial_{[\mu} \lambda^i) \alpha_{-i} \tau_{\nu]}^- \label{gaugetransb} \, ,
\eea
while the remaining fields do not transform under gauge transformations. 
From the previous transformations, we learn that $a_\mu^i$ transforms as a gauge connection, $\alpha_{-}^i$ transforms as a vector and $b_{\mu\nu}$ transforms in a non-covariant way, emulating a Green-Schwarz mechanism. While in the relativistic case this transformation is non-ambiguous and cannot be removed, when one considers the NR expansions presented in this work (based on \cite{EandD}), it is possible to define the field 
\bea
\bar b_{\mu\nu} = b_{\mu\nu} + a_{[\mu| i} \alpha_{-}^i \tau_{|\nu] -}
\label{tildeb}
\eea
which is gauge invariant, i.e. $\delta_{\lambda}\bar b_{\mu \nu}=0$. This is an interesting property of the heterotic supergravity under the NR limit, which will help us to construct a suitable curvature for the Kalb-Ramond field and, also, is potentially telling us that the cancellation of anomalies could be trivial under this limit. If this is the case, one could construct different formulations of heterotic supergravity using arbitrary gauge groups, beyond $SO(32)$ and $E8\times E8$. Finally we observe that the gauge fields do not transform under the boost symmetry, so the field redefinition (\ref{tildeb}) do not deform the boost transformation of the new $\bar b_{\mu \nu}$ field.
 
\subsection{Gauging the covariant derivative}
Considering an arbitrary gauge vector $v^{i}$, upon taking the NR limit, the partial derivative acting on this vector is not covariant. Therefore we need to define a covariant derivative for the gauge symmetry in the following way,
\bea
\nabla_{\mu} v^{i} = \partial_{\mu} v^{i} - f^{i}{}_{jk} a_{\mu}{}^{j} v^{k} \, .
\eea 
We choose $a_{\mu}{}^{j}$ over $\alpha_{-}^{i}$ to construct the covariant derivative due to the transformation property of $a_{\mu}{}^{j}$. In this case, we need a proper connection and $\alpha_{-}^{i}$ transforms as a gauge vector.

Let's start constructing the curvature for the gauge connection $a_\mu^i$, which is given by 
\bea
f_{\mu \nu}{}^{i} = 2 \partial_{[\mu} a_{\nu]}{}^{i} - f^{i}{}_{jk} a_{\mu}{}^{j} a_{\nu}{}^{k} \, ,
\eea
The field $\alpha_{-}^i$ is already covariant, so we do not need to construct its curvature. If we consider $\bar b$ as the fundamental field of the theory, then we define its curvature as  
\bea
\label{newcurva}
\bar h_{\mu \nu \rho} = 3 \partial_{[\mu} \bar b_{\nu\rho]} \, =  h_{\mu \nu \rho} - 3 c^{(g)}_{\mu \nu \rho}.
\eea
From the previous curvature, we can identify the equivalent quantity to the Chern-Simons terms in this limit, 
\bea
c^{(g)}_{\mu \nu \rho} = -\partial_{[\mu}(a_{\nu}^{i} \tau_{\rho]}^{-} \alpha_{-i}).
\eea
The introduction of the $\bar b_{\mu \nu}$ field is a crucial result of this paper. It shows that the gauge Green-Schwarz mechanism is trivial in the NR-limit of the heterotic supergravity. Furthermore, the Chern-Simons terms in form language are given by the exterior derivative of a two form, simplifying the Bianchi identity when we compare to the relativistic case. In section (\ref{Lagrangian}), we will use these curvatures to present the finite NR gauge Lagrangian.

\section{Finite Lagrangian}
\label{Lagrangian}
In this section, we construct all the finite contributions to the heterotic supergravity Lagrangian in covariant form upon taking the NR limit. For readers interested only in the final result, we report the final form of the gauge Lagrangian at the end of this section. 

\subsection{Ungauged Lagrangian}

This Lagrangian contains contributions coming from the Ricci scalar, the dilaton term, and the H-term, given by
\bea
L & = & e e^{-2 \varphi} (R^{(0)} + 4 \partial_{\mu} \varphi \partial^{\mu} \varphi - \frac{1}{12} H^{(0)}_{\mu \nu \rho} H^{(0)}_{\sigma \gamma \epsilon} h^{\mu \sigma} h^{\nu \gamma} h^{\rho \epsilon} \nn \\ && - \frac12 H^{(0)}_{\mu \nu \rho} H^{(2)}_{\sigma \gamma \epsilon} \eta^{a b} \tau^{\mu}{}_{a} \tau^{\sigma}{}_{b} h^{\nu \gamma} h^{\rho \epsilon} - \frac14 H^{(2)}_{\mu \nu \rho} H^{(2)}_{\sigma \gamma \epsilon} \eta^{a b} \eta^{c d} \tau^{\mu}{}_{a} \tau^{\nu}{}_{c} \tau^{\sigma}{}_{b} \tau^{\gamma}{}_{d} h^{\rho \epsilon}) \, ,
\eea
where
\bea
H^{(0)}_{\mu \nu \rho} & = & h_{\mu \nu \rho} = 3 \partial_{[\mu} b_{\nu \rho]} \, , \\
H^{(2)}_{\mu \nu \rho} & = & - 6 c^2 \epsilon_{a b} \nabla_{[\mu} \tau_{\nu}{}^{a} \tau_{\rho]}{}^{b} \, .
\eea
Collecting all the contributions coming from the H-terms, we find the following finite Lagrangian
\bea
L= e e^{-2 \varphi} (R^{(0)} + 4 \partial_{\mu}\varphi  \partial_{\nu} \varphi h^{\mu \nu} + T_{H^2})
\label{Lag1}
\eea
where
\bea
T_{H^2} & = &  2 \eta_{a b} \eta^{c d} \nabla_{[\mu}{\tau_{\nu]}{}^{a}} \nabla_{[\rho}{\tau_{\sigma]}{}^{b}} \tau^{\mu}{}_{c} \tau^{\sigma}{}_{d} h^{\nu \rho} + 2 \epsilon_{a b} \epsilon_{c d} \eta^{a e} \eta^{c f} \nabla_{[\mu}{\tau_{\nu]}{}^{b}} \nabla_{[\rho}{\tau_{\sigma]}{}^{d}} \tau^{\nu}{}_{f} \tau^{\sigma}{}_{e} h^{\mu \rho} \nn \\ && - \frac{1}{12} h_{\mu \nu \rho} h_{\sigma \gamma \epsilon} h^{\mu \sigma} h^{\nu \gamma} h^{\rho \epsilon} + \epsilon_{a b} \eta^{a c} \nabla_{[\mu}{\tau_{\nu]}{}^{b}} h_{\rho \sigma \gamma} \tau^{\sigma}{}_{c} h^{\mu \rho} h^{\nu \gamma}
\label{TH2}
\eea

The Ricci scalar can be easily computed considering the non-relativistic expansions (\ref{metricexpansion})-(\ref{inversemetricexpansion}). The generic form of this object is as follows,
\bea
R = c^{2} R^{(2)} + R^{(0)} + \frac{1}{c^2} R^{-2} + \frac{1}{c^4} R^{-4} \, . \label{Scalar}
\eea
Every contribution in the previous expansion transforms as a scalar, as the reader can easily check (we attach the explicit demonstration for $R^{(0)}$ in the appendix C). The Lagrangian (\ref{Lag1}), together with (\ref{TH2}) and $R^{(0)}$, constitutes the ungauged bosonic part of the heterotic supergravity.

\subsection{Gauged terms}
Now we compute the gauge contributions of the heterotic Lagrangian given by the inclusion of the $\hat A_{\mu}{}^{i}$ field. 

\bea
S_{\rm{het}} = \int d^{10}x \sqrt{-\hat g} e^{-2 \hat \phi} (\hat R + 4 \partial_{\mu} \hat \phi \partial^{\mu} \hat \phi - \frac{1}{12} \bar{H}_{\mu \nu \rho} \bar{H}^{\mu \nu \rho} - \frac14 \hat F_{\mu \nu}{}^{i} \hat F^{\mu \nu}{}_{i}) \, ,
\label{Shet0}
\eea
with
\be
\bar H_{\mu\nu\rho}=3\left(\partial_{[\mu}\hat B_{\nu\rho]}- \hat C_{\mu\nu\rho}^{(g)}\right)\, , \label{barH}
\ee 
\be
\hat C_{\mu\nu\rho}^{(g)}= \hat A^i_{[\mu}\partial_\nu \hat A_{\rho]i}-\frac13 \hat f_{ijk} \hat A_\mu^i \hat A_\nu^j \hat A_\rho^k \, ,
\ee
\bea
\hat F_{\mu \nu}{}^{i} = 2 \partial_{[\mu} \hat A_{\nu]}{}^{i} - \hat f^{i}{}_{jk} \hat A_{\mu}{}^{j} \hat A_{\nu}{}^{k} \, ,
\eea
and $\hat f_{i j k}= c f_{i j k}$. In this case, the Lagrangian also remains finite after considering the limit $c \rightarrow \infty$ and it can be written as 
\bea
L= e e^{-2 \varphi} \Big(R^{(0)} + 4 \partial_{\mu}\varphi  \partial_{\nu} \varphi h^{\mu \nu} + T_
{H^2} + T_{F^2}\Big)
\eea
where
\bea
&& T_{H^2} =  \frac{1}{2} \alpha_{-}^{2}  \eta^{a b} \nabla_{[\mu}{\tau_{\nu]}^{-}} \bar h_{\rho \sigma \gamma} \tau_{\epsilon}^{-} \tau^{\rho}{}_{a} \tau^{\epsilon}{}_{b} h^{\mu \sigma} h^{\nu \gamma} + 2 \alpha_{-}^{2} \epsilon_{a b} \eta^{a c} \eta^{d e} \nabla_{[\mu}{\tau_{\nu]}^{b}} \nabla_{[\rho}{\tau_{\sigma]}^{-}} \tau_{\gamma}^{-} \tau^{\mu}_{d} \tau^{\sigma}_{c} \tau^{\gamma}_{e} h^{\nu \rho} \nn \\ && - \frac{1}{2} \alpha_{-}^{4} \eta^{a b} \eta^{c d} \nabla_{[\mu}{\tau_{\nu]}^{-}} \nabla_{[\rho}{\tau_{\sigma]}^{-}} \tau_{\gamma}^{-} \tau_{\epsilon}^{-} \tau^{\mu}_{a} \tau^{\sigma}_{c} \tau^{\gamma}_{d} \tau^{\epsilon}_{b} h^{\nu \rho} \nn \\ && + 2 \eta_{a b} \eta^{c d} \nabla_{[\mu}{\tau_{\nu]}{}^{a}} \nabla_{[\rho}{\tau_{\sigma]}{}^{b}} \tau^{\mu}{}_{c} \tau^{\sigma}{}_{d} h^{\nu \rho} + 2 \epsilon_{a b} \epsilon_{c d} \eta^{a e} \eta^{c f} \nabla_{[\mu}{\tau_{\nu]}{}^{b}} \nabla_{[\rho}{\tau_{\sigma]}{}^{d}} \tau^{\nu}{}_{f} \tau^{\sigma}{}_{e} h^{\mu \rho} \nn \\ && - \frac{1}{12} \bar h_{\mu \nu \rho} \bar h_{\sigma \gamma \epsilon} h^{\mu \sigma} h^{\nu \gamma} h^{\rho \epsilon} - \epsilon_{a b} \eta^{a c} \nabla_{\mu}{\tau_{\nu}{}^{b}} \tau^{\rho}{}_{c} \bar h_{\rho \sigma \gamma} h^{\mu \sigma} h^{\nu \gamma} 
\label{Hcontributions}
\eea
is the finite contribution coming from the $\bar H^2$-term, with $\alpha_{-}^2=\alpha_{-}^{i} \alpha_{-}^{j} \eta_{i j}$, and the gauge-invariant 3-form is given by 
\bea
\bar h_{\mu \nu \rho} =  h_{\mu \nu \rho} + 3 \partial_{[\mu}(a_{\nu}^{i} \tau_{\rho]}^{-} \alpha_{-i})
\eea

The finite contribution coming from the $\hat F^2$-term is given by
\bea
T_{F^2} & = & - 2 \alpha_{-}^{2} \eta^{a b} \nabla_{[\mu}{\tau_{\nu]}^{-}} \nabla_{[\rho}{\tau_{\sigma]}^{-}} \tau^{\nu}_{a} \tau^{\sigma}_{b} h^{\mu \rho} - 2 \alpha_{-}^{i} \eta^{a b} \eta_{i j} \nabla_{\mu}{\alpha_{-}^{j}} \nabla_{[\nu}{\tau_{\rho]}^{-}} \tau_{\sigma}^{-} \tau^{\rho}_{a} \tau^{\sigma}_{b} h^{\mu \nu} \nn \\ &&  - f_{\mu \nu}^{i} \alpha_{-}^{j} \eta_{i j} \nabla_{\rho}{\tau_{\sigma}^{-}} h^{\mu \rho} h^{\nu \sigma} - \frac{1}{2} \eta^{a b} \eta_{i j} \nabla_{\mu}{\alpha_{-}^{i}} \nabla_{\nu}{\alpha_{-}^{j}} \tau_{\rho}^{-} \tau_{\sigma}^{-} \tau^{\rho}_{a} \tau^{\sigma}_{b} h^{\mu \nu} \, , 
\label{Fcontributions}
\eea
As the reader can easily verify, all the gauge contributions to the heterotic Lagrangian in the NR limit depend on the vector $\alpha_{-}{}^{i}$. 

\subsection{Comparison to the Bergshoeff-Romano formulation}
So far we have computed the transformations and action by considering the alternative expansion
 \bea
 \hat{A}_{\mu}^{i} &= c \ {\tau_\mu}^{-} \alpha_{-}^i + \frac{1}{c} a_\mu^i \, .
\label{comparisonA}
\eea
If we go back to the proposal of Bergshoeff and Romano, we cannot make a direct comparison since the gravitational pair $(h,\tau)_{BR}$ differs form the $(h,\tau)_{LO}$ since they transform differently with respect to gauge transformations. Since $v_{-}{}^{i}$ is boost-invariant and transforms as a vector under gauge transformations, we can study the case $v_{-}{}^{i}=0$, so that all 
\bea
(h,\tau)_{BR} = (h,\tau)_{LO}
\eea
At this point, it is important to observe that taking $\alpha_{-}{}^i=0$ in the prescription used in the previous sections is not equivalent to $v_{-}{}^{i}=0$. In the formalism \cite{BandR} this condition still produces a non-covariant Green-Schwarz mechanism for the $\tilde b$-field, given by
\bea
\delta_{\lambda} \tilde b_{\mu \nu} & = & 2 \tau_{[\mu}{}^{+} e_{\nu]}{}^{a'} (v_{+}{}^i \partial_{a'}\lambda_i - v_{a' i} \partial_{+}\lambda^{i}) - 2 e_{[\mu}{}^{a'} e_{\nu]}{}^{b'} v_{a'}{}^{i} \partial_{b'} \lambda_{i} \nn \\ && - 4 (\tau_{[\mu}{}^{-} \tau_{\nu]}{}^{+} v_{+i} + \tau_{[\mu}{}^{-} e_{\nu]}{}^{a'} v_{a'i} )\partial_{-} \lambda^{i} \, .
\label{GSBR}
\eea
If this transformation could be trivialized by imposing v-field redefinitions, then both setups would be equivalent at the supergravity level. However, the transformation (\ref{GSBR}) is not ambiguous and it cannot be eliminated with a field redefinition. A structural feature of this transformation is that every term has the form $v \times \delta_{\lambda} v$, exactly as happens in the relativistic Green-Schwarz mechanism. 

The previous analysis does not imply that the BR and LO formulations are necessary different. When $v_{-}{}^{i}\neq 0$, one can try to use the anomalous transformation of $\tilde \tau_{\mu}{}^{+}$ to trivialize the Green-Schwarz mechanism but at the moment, is not straightforward to find such field redefinitions. One might try to explore the non-relativistic double field theory in terms of $\hat{\cal H}_{\cal M N}(\hat{g},\hat{B},\hat{A})|_{BR}$ vs $\hat{\cal H}_{\cal M N}(\hat{g},\hat{B},\hat{A})|_{LO} $ and try to relate them by using (generalized) symmetry transformations or field redefinitions at the double geometry. 

At the level of the supergravity action, the advantage of (\ref{comparisonA}) is that one can cleanly construct the action in a manifestly gauge covariant form. Conversely, the highly non-trivial form of (\ref{GSBR}) makes it technically challenging to construct Chern–Simons terms to write the associated action in its covariant form, even when $v_{-}{}^{i}=0$. For interested readers, the explicit finite gauged action of this formulation, in a non-covariant form, is written in the Appendix B.

\section{Discussion}
\label{Conclusions}
In this work, we study the finite form of the heterotic supergravity Lagrangian, extending the formulation of \cite{NSNS}, and providing an alternative way to that of \cite{BandR} of taking this limit. The bosonic part of the Lagrangian of this theory contains two divergent parts given by the Chern-Simons terms and the Yang-Mills Lagrangian which exactly compensate each other, as it happens in bosonic supergravity, giving rise to a finite Lagrangian. The gauge Lagrangian depends on a pair of gauge fields, $a_\mu^i$ and $\alpha_{-}^i$, the former transforming as a gauge connection, so we construct its curvature, and the latter transforms as a gauge vector, so we do not need to construct a curvature. While the formulation presented here is not the first non-relativistic heterotic supergravity in the literature, it is likely that all these approaches are connected by field redefinitions at the double field theory level.

In this work we also compute part of the finite Lagrangian of the formulation given in \cite{BandR} for a particular case where the vielbein is gauge invariant (Appendix B). While there is not yet a proof, a possible way to connect these two formulations of non-relativistic heterotic supergravity is to trivialize the gauge transformation of the vielbein in both theories (without imposing extra constraints), which will need field redefinitions on the b-field and the gauge fields, leading to the same Green-Schwarz mechanism. Particularly, one  needs to match the gauge transformation rules of the gravitational fields associated with the vielbein $(h,\tau)$, since in the Bergshoeff and Romano prescription the $\tau_{\mu}{}^{+}$ transforms in a non-covariant way. In this work we analyze the case $v_{-}{}^{i}=0$, which allows us to eliminate this transformation without imposing field redefinitions ($\delta b_{\mu \nu}$ has a ambiguous Green-Schwarz transformation and $\delta \tilde b_{\mu \nu}$ non-ambiguous Green-Schwarz transformation). In this particular limit ($v_{-}{}^{i}=0$), both prescriptions are different and cannot be related by further field redefinitions. For more general cases where $v_{-}{}^{i}\neq 0$, the compatibility between the two approaches requires the trivialization of 
\bea
\delta \tilde \tau_{\mu}{}^{+} = \frac{v_{-}{}^{i} \partial_{-} \lambda^{i}}{1+v_{+}{}^{i} v_{-i}} \tilde \tau_{\mu}{}^{-} \, .
\eea
The only way to induce a term that transforms proportional to $\partial_{-} \lambda^{i}$ is to consider a field redefinition which contains $\tilde b_{\mu \nu}$ (which is the field whose transformation is proportional to $\partial_{-} \lambda^{i}$), and due to the complexity of the full transformation of the $\tilde b$- field, it is not straightforward to prove the equivalence between the two formalisms. This opens two clear possibilities: 

i) If the theories match using field redefinitions, the results of this paper indicates that the gauge Green-Schwarz transformation can be trivialized and one can impose those field redefinitions to simplify the transformations and use the gauge covariant action presented in this work. Moreover, the cancellation of the Green-Schwarz transformation could indicate that the gauge anomaly could be trivially resolved under the NR limit. We will discuss this point in the outlook, as a promising future direction of work.

 ii) If one can prove that the theories are different, then the results of this paper propose a full consistent alternative approach.

\subsection{Outlook}
This paper opens several promising directions for future work. First, it would be very interesting to study what happens to the equations of motion under the NR limit, especially from the DFT point of view. Even in the bosonic case, these equations are finite by construction (since the generalized metric is finite) and therefore there are no equations missing in the double geometry. However, using the strong constraint to break the duality group could be related to the analysis of the equations of motion performed in \cite{NSNS}. When the analysis is performed directly from the NR supergravity perspective, the Poisson equation is missed and therefore it is not very clear how this equation is lost upon parametrization of the generalized metric. Once this mechanism is understood for the bosonic case, the inclusion of the gauge group should be straightforward. 

Second, a very promising line of research is behind the cancellation of the Green-Schwarz mechanism under the NR expansion shown in this paper. Using the field redefinition given in (\ref{tildeb}), we have shown that the Green-Schwarz mechanism for gauge transformations can be trivialized. The key point here is the presence of the vector $\alpha_{-i}$, instead of the connection $a_{\mu i}$, which allowed us to remove this transformation. This is a strong indication that the anomaly cancellation in the heterotic string could not need a fixed gauge group to work under the NR limit. If this is the case, the present results open the possibility of new families of heterotic strings, considering arbitrary gauge groups. A detailed analysis of the anomaly cancellation is therefore an important step towards the construction of NR heterotic string theory.  

Third, another potential continuation is to include ${\cal N}=1$ supersymmetry in this framework, extending the results in \cite{NRST11}. The inclusion of this symmetry in heterotic DFT requires the use of the generalized vielbein and, therefore, it is not straightforward to construct a finite action in the double geometry. However, once this construction is understood, the inclusion of the gaugino dynamics will be obtained by promoting the duality group from $O(D,D)$ to $O(D,D+N)$. So, this is a worthy program to explore NR heterotic supergravity in the presence of fermions.

Finally, a very ambitious continuation would be the incorporation of $\alpha'$-corrections to both the bosonic and the heterotic supergravity \footnote{See \cite{Electure} for a pedagogical review on this topic and references therein.} in the NR limit. Since the Riemann tensor contains divergences, one should study whether these are controlled by the $\hat H_{\mu \nu \rho}$ (or $\bar H_{\mu \nu \rho}$ in the heterotic case) contributions, as happens at the two-derivative level. The covariant way of writing the Ricci scalar using the covariant derivative of $\tau_{\mu}{}^{a}$ and $\tau^{\mu}{}_{a}$ could be easily extended in order to incorporate the spin connection $\omega_{\mu a b}$, so that the action is also invariant under the $SO(1,1)$ transformations. In this way, the higher-derivative contributions could be written in a more compact form, since we expect large contributions if the curvatures are not chosen in an optimal way. 

While DFT has been proved to be a powerful tool for reading the expansion of the A-field \cite{EandD}, taking the NR limit at the DFT level including higher-derivative terms is probably the best strategy to read these corrections and then parametrize them at the supergravity level. While in this paper we show that the Green-Schwarz mechanism for the gauge symmetry can be trivialized using field redefinitions, it is not clear that the same should happen for the gravitational one when addressing the $\alpha'$-corrections for heterotic supergravity under the non-relativistic limit.  

\subsection*{Acknowledgements}
The author is supported by the SONATA BIS grant 2021/42/E/ST2/00304 from the National Science Centre (NCN), Poland.

\appendix
\section{Details on the cancellation of divergences when $\hat A_{\mu}$ is included}
\label{ap+}
The field strength $\hat F_{\mu \nu}{}^{i}$, after expanding $\hat A_{\mu}{}^{i}$ considering (\ref{Aexpansion}), is given by
\bea
\hat F^{(1)}_{\mu \nu}{}^{i} & = & \alpha_{-}^{i} \partial_{\mu}{\tau_{\nu}^{-}} + \partial_{\mu}{\alpha_{-}^{i}} \tau_{\nu}^{-} - \alpha_{-}^{i} \partial_{\nu}{\tau_{\mu}^{-}} - \partial_{\nu}{\alpha_{-}^{i}} \tau_{\mu}^{-} \, , \\
\hat F^{(0)}_{\mu \nu}{}^{i} & = & + \alpha_{-}^{j} \eta^{i k} \tau_{\mu}^{-} a_{\nu}^{l} f_{j k l} - \alpha_{-}^{j} \eta^{i k} \tau_{\nu}^{-} a_{\mu}^{l} f_{j k l} \, ,
\\ 
\hat F^{(-1)}_{\mu \nu}{}^{i} & = & \partial_{\mu}{a_{\nu}^{i}} - \partial_{\nu}{a_{\mu}^{i}} \, ,  
\\
\hat F^{(-2)}_{\mu \nu}{}^{i} & = & - \eta^{i j} a_{\mu}^{k} a_{\nu}^{l} f_{j k l} \, ,
\eea
where the notation $\hat F^{(i)}_{\mu \nu}$  denotes the terms of order $c^i$ in the expansion. The contraction of this field strength with two inverse metrics, $\hat F^{\mu \nu i}= \hat F_{\rho \sigma}{}^{i} \hat g^{\mu \rho} \hat g^{\nu \sigma}$ yields the expression
\bea
\hat F^{\mu \nu i} = \hat F^{(1)\mu \nu i} + \mathcal{O}\big(\frac{1}{c} \big) \, ,   
\eea
with
\bea
\hat F^{(1)\mu \nu i} = \alpha_{-}^{i} \partial_{\rho}{\tau_{\sigma}^{-}} h^{\mu \rho} h^{\nu \sigma} - \alpha_{-}^{i} \partial_{\rho}{\tau_{\sigma}^{-}} h^{\mu \sigma} h^{\nu \rho} \, .
\eea
As the reader can easily check, the only contraction which produces non-vanishing divergences is given by $\hat F_{(1)\mu \nu}{}^{i} \hat F^{(1)\mu \nu j} \eta_{i j}$, which leads to the result (\ref{YMdiv}). 

Before analyzing the $\bar H$ squared contribution, we recall the form of the Chern-Simons terms,
\be
\bar H_{\mu\nu\rho}=3\left(\partial_{[\mu}\hat B_{\nu\rho]}- \hat C_{\mu\nu\rho}^{(g)}\right)\, , \label{barH}
\ee 
\be
\hat C_{\mu\nu\rho}^{(g)}= \hat A^i_{[\mu}\partial_\nu \hat A_{\rho]i}-\frac13 \hat f_{ijk} \hat A_\mu^i \hat A_\nu^j \hat A_\rho^k \, .
\ee

For the $- \frac{1}{12} \bar H_{\mu \nu \rho} \bar H^{\mu \nu \rho}$ contributions, an important observation is that
\bea
\bar H_{\mu \nu \rho} = \bar H^{(2)}_{\mu \nu \rho} + \bar H_{(0)\mu \nu \rho}  + \mathcal{O}\big(\frac{1}{c}\big) \, , 
\eea
while the same object, when all the indices raised by the inverse metric, has the following form
\bea
\bar H^{\mu \nu \rho} = \bar H^{(0) \mu \nu \rho} + \mathcal{O}\big(\frac{1}{c^2}\big) \, . 
\eea
Thus, we focus on the objects that contribute to divergences,
\bea
\bar H^{(2)}_{\mu \nu \rho} & = & - \epsilon_{a b} \partial_{\mu}{\tau_{\nu}^{a}} \tau_{\rho}^{b}  + \epsilon_{a b} \partial_{\mu}{\tau_{\rho}^{a}} \tau_{\nu}^{b}  - \epsilon_{a b} \partial_{\nu}{\tau_{\rho}^{a}} \tau_{\mu}^{b}  + \epsilon_{a b} \partial_{\nu}{\tau_{\mu}^{a}} \tau_{\rho}^{b}  - \epsilon_{a b} \partial_{\rho}{\tau_{\mu}^{a}} \tau_{\nu}^{b}  \nn \\ && + \epsilon_{a b} \partial_{\rho}{\tau_{\nu}^{a}} \tau_{\mu}^{b}  - \frac12 \alpha_{-}^{i} \alpha_{-}^{j} \eta_{i j} \partial_{\nu}{\tau_{\rho}^{-}} \tau_{\mu}^{-}  - \frac12 \alpha_{-}^{i} \alpha_{-}^{j} \eta_{i j} \partial_{\rho}{\tau_{\mu}^{-}} \tau_{\nu}^{-} \nn \\ && - \frac12 \alpha_{-}^{i} \alpha_{-}^{j} \eta_{i j} \partial_{\mu}{\tau_{\nu}^{-}} \tau_{\rho}^{-} + \frac12 \alpha_{-}^{i} \alpha_{-}^{j} \eta_{i j} \partial_{\rho}{\tau_{\nu}^{-}} \tau_{\mu}^{-} \nn \\ && + \frac12 \alpha_{-}^{i} \alpha_{-}^{j} \eta_{i j} \partial_{\mu}{\tau_{\rho}^{-}} \tau_{\nu}^{-} + \frac12 \alpha_{-}^{i} \alpha_{-}^{j} \eta_{i j} \partial_{\nu}{\tau_{\mu}^{-}} \tau_{\rho}^{-} \, ,  
\eea
\bea
\bar H^{(0)\mu \nu \rho} & = & h_{\sigma \gamma \epsilon} h^{\mu \sigma} h^{\nu \gamma} h^{\rho \epsilon} + \epsilon_{a b} \eta^{a c} \partial_{\sigma}{\tau_{\gamma}^{b}} \tau^{\rho}_{c} h^{\mu \sigma} h^{\nu \gamma} - \epsilon_{a b} \eta^{a c} \partial_{\sigma}{\tau_{\gamma}^{b}} \tau^{\nu}_{c} h^{\mu \sigma} h^{\rho \gamma} \nn \\ && + \epsilon_{a b} \eta^{a c} \partial_{\sigma}{\tau_{\gamma}^{b}} \tau^{\mu}_{c} h^{\nu \sigma} h^{\rho \gamma} - \epsilon_{a b} \eta^{a c} \partial_{\sigma}{\tau_{\gamma}^{b}} \tau^{\rho}_{c} h^{\mu \gamma} h^{\nu \sigma} + \epsilon_{a b} \eta^{a c} \partial_{\sigma}{\tau_{\gamma}^{b}} \tau^{\nu}_{c} h^{\mu \gamma} h^{\rho \sigma} \nn \\ && - \epsilon_{a b} \eta^{a c} \partial_{\sigma}{\tau_{\gamma}^{b}} \tau^{\mu}_{c} h^{\nu \gamma} h^{\rho \sigma} - \frac12 \alpha_{-}^{i} \alpha_{-}^{j} \eta^{a b} \eta_{i j} \partial_{\sigma}{\tau_{\gamma}^{-}} \tau_{\epsilon}^{-} \tau^{\mu}_{a} \tau^{\epsilon}_{b} h^{\nu \sigma} h^{\rho \gamma} \nn \\ && - \frac12 \alpha_{-}^{i} \eta_{i j} \partial_{\sigma}{\tau_{\gamma}^{-}} a_{\epsilon}^{j} h^{\mu \epsilon} h^{\nu \sigma} h^{\rho \gamma} - \frac12 \alpha_{-}^{i} \alpha_{-}^{j} \eta^{a b} \eta_{i j} \partial_{\sigma}{\tau_{\gamma}^{-}} \tau_{\epsilon}^{-} \tau^{\nu}_{a} \tau^{\epsilon}_{b} h^{\mu \gamma} h^{\rho \sigma} \nn \\ && - \frac12 \alpha_{-}^{i} \eta_{i j} \partial_{\sigma}{\tau_{\gamma}^{-}} a_{\epsilon}^{j} h^{\mu \gamma} h^{\nu \epsilon} h^{\rho \sigma} - \frac12 \alpha_{-}^{i} \alpha_{-}^{j} \eta^{a b} \eta_{i j} \partial_{\sigma}{\tau_{\gamma}^{-}} \tau_{\epsilon}^{-} \tau^{\rho}_{a} \tau^{\epsilon}_{b} h^{\mu \sigma} h^{\nu \gamma} \nn \\ && - \frac12 \alpha_{-}^{i} \eta_{i j} \partial_{\sigma}{\tau_{\gamma}^{-}} a_{\epsilon}^{j} h^{\mu \sigma} h^{\nu \gamma} h^{\rho \epsilon} + \frac12 \alpha_{-}^{i} \alpha_{-}^{j} \eta^{a b} \eta_{i j} \partial_{\sigma}{\tau_{\gamma}^{-}} \tau_{\epsilon}^{-} \tau^{\mu}_{a} \tau^{\epsilon}_{b} h^{\nu \gamma} h^{\rho \sigma} \nn \\ && + \frac12 \alpha_{-}^{i} \eta_{i j} \partial_{\sigma}{\tau_{\gamma}^{-}} a_{\epsilon}^{j} h^{\mu \epsilon} h^{\nu \gamma} h^{\rho \sigma} + \frac12 \alpha_{-}^{i} \alpha_{-}^{j} \eta^{a b} \eta_{i j} \partial_{\sigma}{\tau_{\gamma}^{-}} \tau_{\epsilon}^{-} \tau^{\nu}_{a} \tau^{\epsilon}_{b} h^{\mu \sigma} h^{\rho \gamma} \nn \\ && + \frac12 \alpha_{-}^{i} \eta_{i j} \partial_{\sigma}{\tau_{\gamma}^{-}} a_{\epsilon}^{j} h^{\mu \sigma} h^{\nu \epsilon} h^{\rho \gamma} + \frac12 \alpha_{-}^{i} \alpha_{-}^{j} \eta^{a b} \eta_{i j} \partial_{\sigma}{\tau_{\gamma}^{-}} \tau_{\epsilon}^{-} \tau^{\rho}_{a} \tau^{\epsilon}_{b} h^{\mu \gamma} h^{\nu \sigma} \nn \\ && + \frac12 \alpha_{-}^{i} \eta_{i j} \partial_{\sigma}{\tau_{\gamma}^{-}} a_{\epsilon}^{j} h^{\mu \gamma} h^{\nu \sigma} h^{\rho \epsilon} \, .
\eea
The multiplication of the previous two quantities leads directly to the result (\ref{barH2cancel}) after imposing the Newton-Cartan relations (\ref{NCrelations}).

\section{The Bergshoeff-Romano action}
The finite action from the construction given in \cite{BandR} in the case $v_{-}{}^{i}=0$ is given by
\bea
L= e e^{-2 \varphi} \Big(L_{\textrm ungauged} + T_{H^2} + T_{F^2}\Big)
\label{BandR}
\eea
where the gauged contributions of the Lagrangian are given by (we use A,B... for the transversal indices and tilde fields to differentiate them),
\bea
&& T_{H^2} =  \frac{1}{2} \alpha_{+}^{i} \eta_{i j} \partial_{\mu}{\tilde \tau_{\nu}^{+}} \partial_{\rho}{\tilde b_{\sigma \gamma}} a_{A}^{j} e_{\epsilon}^{A} h^{\mu \sigma} h^{\nu \gamma} h^{\rho \epsilon} + \frac{1}{2} \eta_{i j} \partial_{\mu}{\tilde b_{\nu \rho}} \partial_{\sigma}{e_{\gamma}^{A}} a_{A}^{i} a_{B}^{j} e_{\epsilon}^{B} h^{\mu \epsilon} h^{\nu \sigma} h^{\rho \gamma} \nn \\ && + \frac{1}{2} \eta_{i j} \partial_{\mu}{a_{A}^{i}} \partial_{\nu}{\tilde b_{\rho \sigma}} a_{B}^{j} e_{\gamma}^{A} e_{\epsilon}^{B} h^{\mu \rho} h^{\nu \epsilon} h^{\sigma \gamma} - \frac{1}{2} \alpha_{+}^{i} \eta_{i j} \partial_{\mu}{\tilde \tau_{\nu}^{+}} \partial_{\rho}{\tilde b_{\sigma \gamma}} a_{A}^{j} e_{\epsilon}^{A} h^{\mu \sigma} h^{\nu \rho} h^{\gamma \epsilon} \nn \\ && - \frac{1}{2} \eta_{i j} \partial_{\mu}{\tilde b_{\nu \rho}} \partial_{\sigma}{e_{\gamma}^{A}} a_{A}^{i} a_{B}^{j} e_{\epsilon}^{B} h^{\mu \gamma} h^{\nu \sigma} h^{\rho \epsilon} - \frac{1}{2} \eta_{i j} \partial_{\mu}{a_{A}^{i}} \partial_{\nu}{\tilde b_{\rho \sigma}} a_{B}^{j} e_{\gamma}^{A} e_{\epsilon}^{B} h^{\mu \rho} h^{\nu \gamma} h^{\sigma \epsilon} \nn \\ 
&& 
+ \frac{1}{2} \alpha_{+}^{i} \eta_{i j} \partial_{\mu}{\tilde \tau_{\nu}^{+}} \partial_{\rho}{\tilde b_{\sigma \gamma}} a_{A}^{j} e_{\epsilon}^{A} h^{\mu \rho} h^{\nu \sigma} h^{\gamma \epsilon} + \frac{1}{2} \eta_{i j} \partial_{\mu}{\tilde b_{\nu \rho}} \partial_{\sigma}{e_{\gamma}^{A}} a_{A}^{i} a_{B}^{j} e_{\epsilon}^{B} h^{\mu \sigma} h^{\nu \gamma} h^{\rho \epsilon} \nn \\ && + \frac{1}{2} \eta_{i j} \partial_{\mu}{a_{A}^{i}} \partial_{\nu}{\tilde b_{\rho \sigma}} a_{B}^{j} e_{\gamma}^{A} e_{\epsilon}^{B} h^{\mu \nu} h^{\rho \gamma} h^{\sigma \epsilon} - \frac{1}{2} \partial_{\mu}{\tilde b_{\nu \rho}} a_{A}^{i} a_{B}^{j} a_{C}^{k} e_{\sigma}^{A} e_{\gamma}^{B} e_{\epsilon}^{C} f_{i j k} h^{\mu \sigma} h^{\nu \gamma} h^{\rho \epsilon} \nn \\ && - \frac{1}{2} \alpha_{+}^{i} \epsilon_{a b} \eta^{a c} \eta_{i j} \partial_{\mu}{\tilde \tau_{\nu}^{b}} \partial_{\rho}{\tilde \tau_{\sigma}^{+}} \tilde \tau^{\sigma}_{c} a_{A}^{j} e_{\gamma}^{A} h^{\mu \gamma} h^{\nu \rho} - \frac{1}{2} \epsilon_{a b} \eta^{a c} \eta_{i j} \partial_{\mu}{\alpha_{+}^{i}} \partial_{\nu}{\tilde \tau_{\rho}^{b}} \tilde \tau_{\sigma}^{+} \tilde \tau^{\sigma}_{c} a_{A}^{j} e_{\gamma}^{A} h^{\mu \rho} h^{\nu \gamma} \nn \\ && - \frac{1}{2} \epsilon_{a b} \eta^{a c} \eta_{i j} \partial_{\mu}{\tilde \tau_{\nu}^{b}} \partial_{\rho}{e_{\sigma}^{A}} \tilde \tau^{\sigma}_{c} a_{A}^{i} a_{B}^{j} e_{\gamma}^{B} h^{\mu \gamma} h^{\nu \rho} - \frac{1}{2} \alpha_{+}^{i} \epsilon_{a b} \eta^{a c} \eta_{i j} \partial_{\mu}{\tilde \tau_{\nu}^{b}} \partial_{\rho}{\tilde \tau_{\sigma}^{+}} \tilde \tau^{\rho}_{c} a_{A}^{j} e_{\gamma}^{A} h^{\mu \sigma} h^{\nu \gamma} \nn \\ && - \frac{1}{2} \epsilon_{a b} \eta^{a c} \eta_{i j} \partial_{\mu}{\tilde \tau_{\nu}^{b}} \partial_{\rho}{e_{\sigma}^{A}} \tilde \tau^{\rho}_{c} a_{A}^{i} a_{B}^{j} e_{\gamma}^{B} h^{\mu \sigma} h^{\nu \gamma} - \frac{1}{2} \epsilon_{a b} \eta^{a c} \eta_{i j} \partial_{\mu}{\tilde \tau_{\nu}^{b}} \partial_{\rho}{a_{A}^{i}} \tilde \tau^{\rho}_{c} a_{B}^{j} e_{\sigma}^{A} e_{\gamma}^{B} h^{\mu \sigma} h^{\nu \gamma} \nn \\ && - \frac{1}{2} \alpha_{+}^{i} \alpha_{+}^{j} \epsilon_{a b} \eta^{a c} \eta_{i j} \partial_{\mu}{\tilde \tau_{\nu}^{b}} \partial_{\rho}{\tilde \tau_{\sigma}^{+}} \tilde \tau_{\gamma}^{+} \tilde \tau^{\gamma}_{c} h^{\mu \rho} h^{\nu \sigma} - \frac{1}{2} \alpha_{+}^{i} \epsilon_{a b} \eta^{a c} \eta_{i j} \partial_{\mu}{\tilde \tau_{\nu}^{b}} \partial_{\rho}{e_{\sigma}^{A}} \tilde \tau_{\gamma}^{+} \tilde \tau^{\gamma}_{c} a_{A}^{j} h^{\mu \rho} h^{\nu \sigma} \nn \\ && - \frac{1}{2} \alpha_{+}^{i} \epsilon_{a b} \eta^{a c} \eta_{i j} \partial_{\mu}{\tilde \tau_{\nu}^{b}} \partial_{\rho}{a_{A}^{j}} \tilde \tau_{\sigma}^{+} \tilde \tau^{\sigma}_{c} e_{\gamma}^{A} h^{\mu \rho} h^{\nu \gamma} + \frac{1}{2} \alpha_{+}^{i} \epsilon_{a b} \eta^{a c} \eta_{i j} \partial_{\mu}{\tilde \tau_{\nu}^{b}} \partial_{\rho}{\tilde \tau_{\sigma}^{+}} \tilde \tau^{\rho}_{c} a_{A}^{j} e_{\gamma}^{A} h^{\mu \gamma} h^{\nu \sigma} \nn \\ && + \frac{1}{2} \epsilon_{a b} \eta^{a c} \eta_{i j} \partial_{\mu}{\tilde \tau_{\nu}^{b}} \partial_{\rho}{e_{\sigma}^{A}} \tilde \tau^{\rho}_{c} a_{A}^{i} a_{B}^{j} e_{\gamma}^{B} h^{\mu \gamma} h^{\nu \sigma} + \frac{1}{2} \epsilon_{a b} \eta^{a c} \eta_{i j} \partial_{\mu}{\tilde \tau_{\nu}^{b}} \partial_{\rho}{a_{A}^{i}} \tilde \tau^{\rho}_{c} a_{B}^{j} e_{\sigma}^{A} e_{\gamma}^{B} h^{\mu \gamma} h^{\nu \sigma} \nn \\ && + \frac{1}{2} \alpha_{+}^{i} \epsilon_{a b} \eta^{a c} \eta_{i j} \partial_{\mu}{\tilde \tau_{\nu}^{b}} \partial_{\rho}{\tilde \tau_{\sigma}^{+}} \tilde \tau^{\sigma}_{c} a_{A}^{j} e_{\gamma}^{A} h^{\mu \rho} h^{\nu \gamma} + \frac{1}{2} \epsilon_{a b} \eta^{a c} \eta_{i j} \partial_{\mu}{\alpha_{+}^{i}} \partial_{\nu}{\tilde \tau_{\rho}^{b}} \tilde \tau_{\sigma}^{+} \tilde \tau^{\sigma}_{c} a_{A}^{j} e_{\gamma}^{A} h^{\mu \nu} h^{\rho \gamma} \nn \\ && + \frac{1}{2} \epsilon_{a b} \eta^{a c} \eta_{i j} \partial_{\mu}{\tilde \tau_{\nu}^{b}} \partial_{\rho}{e_{\sigma}^{A}} \tilde \tau^{\sigma}_{c} a_{A}^{i} a_{B}^{j} e_{\gamma}^{B} h^{\mu \rho} h^{\nu \gamma} + \frac{1}{2} \alpha_{+}^{i} \alpha_{+}^{j} \epsilon_{a b} \eta^{a c} \eta_{i j} \partial_{\mu}{\tilde \tau_{\nu}^{b}} \partial_{\rho}{\tilde \tau_{\sigma}^{+}} \tilde \tau_{\gamma}^{+} \tilde \tau^{\gamma}_{c} h^{\mu \sigma} h^{\nu \rho} \nn \\ && + \frac{1}{2} \alpha_{+}^{i} \epsilon_{a b} \eta^{a c} \eta_{i j} \partial_{\mu}{\tilde \tau_{\nu}^{b}} \partial_{\rho}{e_{\sigma}^{A}} \tilde \tau_{\gamma}^{+} \tilde \tau^{\gamma}_{c} a_{A}^{j} h^{\mu \sigma} h^{\nu \rho} + \frac{1}{2} \alpha_{+}^{i} \epsilon_{a b} \eta^{a c} \eta_{i j} \partial_{\mu}{\tilde \tau_{\nu}^{b}} \partial_{\rho}{a_{A}^{j}} \tilde \tau_{\sigma}^{+} \tilde \tau^{\sigma}_{c} e_{\gamma}^{A} h^{\mu \gamma} h^{\nu \rho} \nn \\ && + \alpha_{+}^{i} \epsilon_{a b} \eta^{a c} \partial_{\mu}{\tilde \tau_{\nu}^{b}} \tilde \tau_{\rho}^{+} \tilde \tau^{\rho}_{c} a_{A}^{j} a_{B}^{k} e_{\sigma}^{A} e_{\gamma}^{B} f_{i j k} h^{\mu \sigma} h^{\nu \gamma} - \frac{1}{8} \alpha_{+}^{i} \alpha_{+}^{j} \eta_{i k} \eta_{j i1} \partial_{\mu}{\tilde \tau_{\nu}^{+}} \partial_{\rho}{\tilde \tau_{\sigma}^{+}} a_{A}^{k} a_{B}^{i1} e_{\gamma}^{A} e_{\epsilon}^{B} h^{\mu \rho} h^{\nu \sigma} h^{\gamma \epsilon} \nn \\ && - \frac{1}{4} \alpha_{+}^{i} \eta_{i j} \eta_{k i1} \partial_{\mu}{\tilde \tau_{\nu}^{+}} \partial_{\rho}{e_{\sigma}^{A}} a_{A}^{k} a_{B}^{j} a_{C}^{i1} e_{\gamma}^{B} e_{\epsilon}^{C} h^{\mu \rho} h^{\nu \sigma} h^{\gamma \epsilon} - \frac{1}{4} \alpha_{+}^{i} \eta_{i j} \eta_{k i1} \partial_{\mu}{\tilde \tau_{\nu}^{+}} \partial_{\rho}{a_{A}^{k}} a_{B}^{j} a_{C}^{i1} e_{\sigma}^{A} e_{\gamma}^{B} e_{\epsilon}^{C} h^{\mu \rho} h^{\nu \sigma} h^{\gamma \epsilon} \nn \\ && 
 - \frac{1}{4} \alpha_{+}^{i} \alpha_{+}^{j} \eta_{i k} \eta_{j i1} \partial_{\mu}{\tilde \tau_{\nu}^{+}} \partial_{\rho}{\tilde \tau_{\sigma}^{+}} a_{A}^{k} a_{B}^{i1} e_{\gamma}^{A} e_{\epsilon}^{B} h^{\mu \sigma} h^{\nu \epsilon} h^{\rho \gamma} - \frac{1}{4} \alpha_{+}^{i} \eta_{i j} \eta_{k i1} \partial_{\mu}{\tilde \tau_{\nu}^{+}} \partial_{\rho}{e_{\sigma}^{A}} a_{A}^{k} a_{B}^{j} a_{C}^{i1} e_{\gamma}^{B} e_{\epsilon}^{C} h^{\mu \epsilon} h^{\nu \rho} h^{\sigma \gamma} \nn \\ && - \frac{1}{4} \alpha_{+}^{i} \eta_{i j} \eta_{k i1} \partial_{\mu}{\tilde \tau_{\nu}^{+}} \partial_{\rho}{a_{A}^{k}} a_{B}^{j} a_{C}^{i1} e_{\sigma}^{A} e_{\gamma}^{B} e_{\epsilon}^{C} h^{\mu \epsilon} h^{\nu \rho} h^{\sigma \gamma} - \frac{1}{4} \alpha_{+}^{i} \eta_{i j} \eta_{k i1} \partial_{\mu}{\tilde \tau_{\nu}^{+}} \partial_{\rho}{e_{\sigma}^{A}} a_{A}^{k} a_{B}^{j} a_{C}^{i1} e_{\gamma}^{B} e_{\epsilon}^{C} h^{\mu \sigma} h^{\nu \epsilon} h^{\rho \gamma} \nn
\eea
\bea
&& - \frac{1}{4} \alpha_{+}^{i} \eta_{i j} \eta_{k i1} \partial_{\mu}{\tilde \tau_{\nu}^{+}} \partial_{\rho}{a_{A}^{k}} a_{B}^{j} a_{C}^{i1} e_{\sigma}^{A} e_{\gamma}^{B} e_{\epsilon}^{C} h^{\mu \sigma} h^{\nu \epsilon} h^{\rho \gamma} + \frac{1}{8} \alpha_{+}^{i} \alpha_{+}^{j} \eta_{i k} \eta_{j i1} \partial_{\mu}{\tilde \tau_{\nu}^{+}} \partial_{\rho}{\tilde \tau_{\sigma}^{+}} a_{A}^{k} a_{B}^{i1} e_{\gamma}^{A} e_{\epsilon}^{B} h^{\mu \sigma} h^{\nu \rho} h^{\gamma \epsilon} \nn \\ &&
+ \frac{1}{4} \alpha_{+}^{i} \eta_{i j} \eta_{k i1} \partial_{\mu}{\tilde \tau_{\nu}^{+}} \partial_{\rho}{e_{\sigma}^{A}} a_{A}^{k} a_{B}^{j} a_{C}^{i1} e_{\gamma}^{B} e_{\epsilon}^{C} h^{\mu \sigma} h^{\nu \rho} h^{\gamma \epsilon} + \frac{1}{4} \alpha_{+}^{i} \eta_{i j} \eta_{k i1} \partial_{\mu}{\tilde \tau_{\nu}^{+}} \partial_{\rho}{a_{A}^{k}} a_{B}^{j} a_{C}^{i1} e_{\sigma}^{A} e_{\gamma}^{B} e_{\epsilon}^{C} h^{\mu \sigma} h^{\nu \rho} h^{\gamma \epsilon} \nn \\ && + \frac{1}{8} \alpha_{+}^{i} \alpha_{+}^{j} \eta_{i k} \eta_{j i1} \partial_{\mu}{\tilde \tau_{\nu}^{+}} \partial_{\rho}{\tilde \tau_{\sigma}^{+}} a_{A}^{k} a_{B}^{i1} e_{\gamma}^{A} e_{\epsilon}^{B} h^{\mu \epsilon} h^{\nu \sigma} h^{\rho \gamma} + \frac{1}{4} \alpha_{+}^{i} \eta_{i j} \eta_{k i1} \partial_{\mu}{\tilde \tau_{\nu}^{+}} \partial_{\rho}{e_{\sigma}^{A}} a_{A}^{k} a_{B}^{j} a_{C}^{i1} e_{\gamma}^{B} e_{\epsilon}^{C} h^{\mu \epsilon} h^{\nu \sigma} h^{\rho \gamma} \nn \\ &&
+ \frac{1}{4} \alpha_{+}^{i} \eta_{i j} \eta_{k i1} \partial_{\mu}{\tilde \tau_{\nu}^{+}} \partial_{\rho}{a_{A}^{k}} a_{B}^{j} a_{C}^{i1} e_{\sigma}^{A} e_{\gamma}^{B} e_{\epsilon}^{C} h^{\mu \epsilon} h^{\nu \sigma} h^{\rho \gamma} + \frac{1}{8} \alpha_{+}^{i} \alpha_{+}^{j} \eta_{i k} \eta_{j i1} \partial_{\mu}{\tilde \tau_{\nu}^{+}} \partial_{\rho}{\tilde \tau_{\sigma}^{+}} a_{A}^{k} a_{B}^{i1} e_{\gamma}^{A} e_{\epsilon}^{B} h^{\mu \rho} h^{\nu \epsilon} h^{\sigma \gamma}  \nn \\
&& + \frac{1}{4} \alpha_{+}^{i} \eta_{i j} \eta_{k i1} \partial_{\mu}{\tilde \tau_{\nu}^{+}} \partial_{\rho}{e_{\sigma}^{A}} a_{A}^{k} a_{B}^{j} a_{C}^{i1} e_{\gamma}^{B} e_{\epsilon}^{C} h^{\mu \rho} h^{\nu \epsilon} h^{\sigma \gamma} + \frac{1}{4} \alpha_{+}^{i} \eta_{i j} \eta_{k i1} \partial_{\mu}{\tilde \tau_{\nu}^{+}} \partial_{\rho}{a_{A}^{k}} a_{B}^{j} a_{C}^{i1} e_{\sigma}^{A} e_{\gamma}^{B} e_{\epsilon}^{C} h^{\mu \rho} h^{\nu \epsilon} h^{\sigma \gamma} \nn \\ && + \frac{1}{2} \alpha_{+}^{i} \eta_{i j} \partial_{\mu}{\tilde \tau_{\nu}^{+}} a_{A}^{j} a_{B}^{k} a_{C}^{i1} a_{D}^{i2} e_{\rho}^{A} e_{\sigma}^{B} e_{\gamma}^{C} e_{\epsilon}^{D} f_{k i1 i2} h^{\mu \sigma} h^{\nu \gamma} h^{\rho \epsilon} - \frac{1}{8} \eta_{i j} \eta_{k i1} \partial_{\mu}{e_{\nu}^{A}} \partial_{\rho}{e_{\sigma}^{B}} a_{A}^{i} a_{B}^{k} a_{C}^{j} a_{D}^{i1} e_{\gamma}^{C} e_{\epsilon}^{D} h^{\mu \rho} h^{\nu \sigma} h^{\gamma \epsilon} \nn \\ && 
- \frac{1}{4} \eta_{i j} \eta_{k i1} \partial_{\mu}{a_{A}^{i}} \partial_{\nu}{e_{\rho}^{B}} a_{B}^{k} a_{C}^{j} a_{D}^{i1} e_{\sigma}^{A} e_{\gamma}^{C} e_{\epsilon}^{D} h^{\mu \nu} h^{\rho \sigma} h^{\gamma \epsilon} - \frac{1}{4} \eta_{i j} \eta_{k i1} \partial_{\mu}{e_{\nu}^{A}} \partial_{\rho}{e_{\sigma}^{B}} a_{A}^{i} a_{B}^{k} a_{C}^{j} a_{D}^{i1} e_{\gamma}^{C} e_{\epsilon}^{D} h^{\mu \sigma} h^{\nu \epsilon} h^{\rho \gamma} \nn \\ && 
- \frac{1}{4} \eta_{i j} \eta_{k i1} \partial_{\mu}{a_{A}^{i}} \partial_{\nu}{e_{\rho}^{B}} a_{B}^{k} a_{C}^{j} a_{D}^{i1} e_{\sigma}^{A} e_{\gamma}^{C} e_{\epsilon}^{D} h^{\mu \rho} h^{\nu \gamma} h^{\sigma \epsilon} - \frac{1}{4} \eta_{i j} \eta_{k i1} \partial_{\mu}{a_{A}^{i}} \partial_{\nu}{e_{\rho}^{B}} a_{B}^{k} a_{C}^{j} a_{D}^{i1} e_{\sigma}^{A} e_{\gamma}^{C} e_{\epsilon}^{D} h^{\mu \epsilon} h^{\nu \sigma} h^{\rho \gamma}
\nn \\ && + \frac{1}{8} \eta_{i j} \eta_{k i1} \partial_{\mu}{e_{\nu}^{A}} \partial_{\rho}{e_{\sigma}^{B}} a_{A}^{i} a_{B}^{k} a_{C}^{j} a_{D}^{i1} e_{\gamma}^{C} e_{\epsilon}^{D} h^{\mu \sigma} h^{\nu \rho} h^{\gamma \epsilon} + \frac{1}{4} \eta_{i j} \eta_{k i1} \partial_{\mu}{a_{A}^{i}} \partial_{\nu}{e_{\rho}^{B}} a_{B}^{k} a_{C}^{j} a_{D}^{i1} e_{\sigma}^{A} e_{\gamma}^{C} e_{\epsilon}^{D} h^{\mu \rho} h^{\nu \sigma} h^{\gamma \epsilon} \nn \\ 
&& + \frac{1}{8} \eta_{i j} \eta_{k i1} \partial_{\mu}{e_{\nu}^{A}} \partial_{\rho}{e_{\sigma}^{B}} a_{A}^{i} a_{B}^{k} a_{C}^{j} a_{D}^{i1} e_{\gamma}^{C} e_{\epsilon}^{D} h^{\mu \epsilon} h^{\nu \sigma} h^{\rho \gamma} + \frac{1}{4} \eta_{i j} \eta_{k i1} \partial_{\mu}{a_{A}^{i}} \partial_{\nu}{e_{\rho}^{B}} a_{B}^{k} a_{C}^{j} a_{D}^{i1} e_{\sigma}^{A} e_{\gamma}^{C} e_{\epsilon}^{D} h^{\mu \epsilon} h^{\nu \gamma} h^{\rho \sigma} \nn \\ && + \frac{1}{8} \eta_{i j} \eta_{k i1} \partial_{\mu}{e_{\nu}^{A}} \partial_{\rho}{e_{\sigma}^{B}} a_{A}^{i} a_{B}^{k} a_{C}^{j} a_{D}^{i1} e_{\gamma}^{C} e_{\epsilon}^{D} h^{\mu \rho} h^{\nu \epsilon} h^{\sigma \gamma} + \frac{1}{4} \eta_{i j} \eta_{k i1} \partial_{\mu}{a_{A}^{i}} \partial_{\nu}{e_{\rho}^{B}} a_{B}^{k} a_{C}^{j} a_{D}^{i1} e_{\sigma}^{A} e_{\gamma}^{C} e_{\epsilon}^{D} h^{\mu \nu} h^{\rho \gamma} h^{\sigma \epsilon} \nn \\ && + \frac{1}{2} \eta_{i j} \partial_{\mu}{e_{\nu}^{A}} a_{A}^{i} a_{B}^{j} a_{C}^{k} a_{D}^{i1} a_{E}^{i2} e_{\rho}^{B} e_{\sigma}^{C} e_{\gamma}^{D} e_{\epsilon}^{E} f_{k i1 i2} h^{\mu \sigma} h^{\nu \gamma} h^{\rho \epsilon} - \frac{1}{8} \eta_{i j} \eta_{k i1} \partial_{\mu}{a_{A}^{i}} \partial_{\nu}{a_{B}^{k}} a_{C}^{j} a_{D}^{i1} e_{\rho}^{A} e_{\sigma}^{B} e_{\gamma}^{C} e_{\epsilon}^{D} h^{\mu \nu} h^{\rho \sigma} h^{\gamma \epsilon}
\nn \\ && - \frac{1}{4} \eta_{i j} \eta_{k i1} \partial_{\mu}{a_{A}^{i}} \partial_{\nu}{a_{B}^{k}} a_{C}^{j} a_{D}^{i1} e_{\rho}^{A} e_{\sigma}^{B} e_{\gamma}^{C} e_{\epsilon}^{D} h^{\mu \sigma} h^{\nu \gamma} h^{\rho \epsilon} + \frac{1}{8} \eta_{i j} \eta_{k i1} \partial_{\mu}{a_{A}^{i}} \partial_{\nu}{a_{B}^{k}} a_{C}^{j} a_{D}^{i1} e_{\rho}^{A} e_{\sigma}^{B} e_{\gamma}^{C} e_{\epsilon}^{D} h^{\mu \sigma} h^{\nu \rho} h^{\gamma \epsilon} \nn \\ && + \frac{1}{8} \eta_{i j} \eta_{k i1} \partial_{\mu}{a_{A}^{i}} \partial_{\nu}{a_{B}^{k}} a_{C}^{j} a_{D}^{i1} e_{\rho}^{A} e_{\sigma}^{B} e_{\gamma}^{C} e_{\epsilon}^{D} h^{\mu \epsilon} h^{\nu \gamma} h^{\rho \sigma} + \frac{1}{8} \eta_{i j} \eta_{k i1} \partial_{\mu}{a_{A}^{i}} \partial_{\nu}{a_{B}^{k}} a_{C}^{j} a_{D}^{i1} e_{\rho}^{A} e_{\sigma}^{B} e_{\gamma}^{C} e_{\epsilon}^{D} h^{\mu \nu} h^{\rho \epsilon} h^{\sigma \gamma}
\nn \\ && + \frac{1}{2} \eta_{i j} \partial_{\mu}{a_{A}^{i}} a_{B}^{j} a_{C}^{k} a_{D}^{i1} a_{E}^{i2} e_{\nu}^{A} e_{\rho}^{B} e_{\sigma}^{C} e_{\gamma}^{D} e_{\epsilon}^{E} f_{k i1 i2} h^{\mu \sigma} h^{\nu \gamma} h^{\rho \epsilon} \nn \\ && - \frac{1}{12} a_{A}^{i} a_{B}^{j} a_{C}^{k} a_{D}^{i1} a_{E}^{i2} a_{F}^{i3} e_{\mu}^{A} e_{\nu}^{B} e_{\rho}^{C} e_{\sigma}^{D} e_{\gamma}^{E} e_{\epsilon}^{F} f_{i j k} f_{i1 i2 i3} h^{\mu \sigma} h^{\nu \gamma} h^{\rho \epsilon} \, , 
\eea
and
\bea
T_
{F^2} & = & - \frac{1}{2} \alpha_{+}^{i} \alpha_{+}^{j} \eta_{i j} \partial_{\mu}{\tilde \tau_{\nu}^{+}} \partial_{\rho}{\tilde \tau_{\sigma}^{+}} h^{\mu \rho} h^{\nu \sigma} - \alpha_{+}^{i} \eta_{i j} \partial_{\mu}{\tilde \tau_{\nu}^{+}} \partial_{\rho}{e_{\sigma}^{A}} a_{A}^{j} h^{\mu \rho} h^{\nu \sigma} - \alpha_{+}^{i} \eta_{i j} \partial_{\mu}{\tilde \tau_{\nu}^{+}} \partial_{\rho}{a_{A}^{j}} e_{\sigma}^{A} h^{\mu \rho} h^{\nu \sigma} \nn \\ && + \frac{1}{2} \alpha_{+}^{i} \alpha_{+}^{j} \eta_{i j} \partial_{\mu}{\tilde \tau_{\nu}^{+}} \partial_{\rho}{\tilde \tau_{\sigma}^{+}} h^{\mu \sigma} h^{\nu \rho} + \alpha_{+}^{i} \eta_{i j} \partial_{\mu}{\tilde \tau_{\nu}^{+}} \partial_{\rho}{e_{\sigma}^{A}} a_{A}^{j} h^{\mu \sigma} h^{\nu \rho} + \alpha_{+}^{i} \eta_{i j} \partial_{\mu}{\tilde \tau_{\nu}^{+}} \partial_{\rho}{a_{A}^{j}} e_{\sigma}^{A} h^{\mu \sigma} h^{\nu \rho} \nn \\ && + \alpha_{+}^{i} \partial_{\mu}{\tilde \tau_{\nu}^{+}} a_{A}^{j} a_{B}^{k} e_{\rho}^{A} e_{\sigma}^{B} f_{i j k} h^{\mu \rho} h^{\nu \sigma} - \frac{1}{2} \eta_{i j} \partial_{\mu}{e_{\nu}^{A}} \partial_{\rho}{e_{\sigma}^{B}} a_{A}^{i} a_{B}^{j} h^{\mu \rho} h^{\nu \sigma} - \eta_{i j} \partial_{\mu}{a_{A}^{i}} \partial_{\nu}{e_{\rho}^{B}} a_{B}^{j} e_{\sigma}^{A} h^{\mu \nu} h^{\rho \sigma} \nn \\ && + \frac{1}{2} \eta_{i j} \partial_{\mu}{e_{\nu}^{A}} \partial_{\rho}{e_{\sigma}^{B}} a_{A}^{i} a_{B}^{j} h^{\mu \sigma} h^{\nu \rho} + \eta_{i j} \partial_{\mu}{a_{A}^{i}} \partial_{\nu}{e_{\rho}^{B}} a_{B}^{j} e_{\sigma}^{A} h^{\mu \rho} h^{\nu \sigma} + \partial_{\mu}{e_{\nu}^{A}} a_{A}^{i} a_{B}^{j} a_{C}^{k} e_{\rho}^{B} e_{\sigma}^{C} f_{i j k} h^{\mu \rho} h^{\nu \sigma} \nn \\ && - \frac{1}{2} \eta_{i j} \partial_{\mu}{a_{A}^{i}} \partial_{\nu}{a_{B}^{j}} e_{\rho}^{A} e_{\sigma}^{B} h^{\mu \nu} h^{\rho \sigma} + \frac{1}{2} \eta_{i j} \partial_{\mu}{a_{A}^{i}} \partial_{\nu}{a_{B}^{j}} e_{\rho}^{A} e_{\sigma}^{B} h^{\mu \sigma} h^{\nu \rho} + \partial_{\mu}{a_{A}^{i}} a_{B}^{j} a_{C}^{k} e_{\nu}^{A} e_{\rho}^{B} e_{\sigma}^{C} f_{i j k} h^{\mu \rho} h^{\nu \sigma} \nn \\ && - \frac{1}{4} \eta^{i j} a_{A}^{k} a_{B}^{i1} a_{C}^{i2} a_{D}^{i3} e_{\mu}^{A} e_{\nu}^{B} e_{\rho}^{C} e_{\sigma}^{D} f_{i k i1} f_{j i2 i3} h^{\mu \rho} h^{\nu \sigma} \, ,
\eea
where $a_{A}^{i} = v_{A}^{i}$ and $\alpha_{+}^{i}=v_{+}^{i}$ transform both as gauge connections. The relation between this construction and the worldsheet formalism of the non-relativistic heterotic supergravity was recently studied in \cite{newB}. 

\section{Demonstration of covariance of $R^{(0)}$}
The difference between the variation of $R^{0}$ using the infinitesimal diffeomorphism transformations,
\bea
\delta_{\xi} \tau_{\mu}{}^{a} & = &  \xi^{\rho} \partial_{\rho}{\tau_{\mu}{}^{a}} + \partial_{\mu}{\xi^{\rho}} \tau_{\rho}{}^{a} \, , \nn \\
\delta_{\xi} \tau^{\mu}{}_{a} & = & \xi^{\rho} \partial_{\rho}{\tau^{\mu}{}_{a}} - \partial_{\rho}{\xi^{\mu}} \tau^{\rho}{}_{a} \, , \nn \\
\delta_{\xi} h^{\mu \nu} & = & \xi^{\rho} \partial_{\rho}{h^{\mu \nu}} – 2 \partial_{\rho}{\xi^{(\mu}} h^{ \nu) \rho} \, , \nn \\ 
\delta_{\xi} h_{\mu \nu} & = & \xi^{\rho} \partial_{\rho}{h_{\mu \nu}} + 2 \partial_{(\mu}{\xi^{\rho}} h_{\nu) \rho} \, ,
\eea
and $\xi^{\rho} \partial_{\rho} R^{(0)}$ (scalar transformation) gives
\bea
&& 2 \partial_{\mu}{h_{\nu \rho}} \partial_{\sigma \gamma}{\xi^{\epsilon}} h_{\epsilon \delta} h^{\mu \nu} h^{\rho \delta} h^{\sigma \gamma} + 2 \partial_{\mu}{\tau_{\nu}^{a}} \partial_{\rho \sigma}{\xi^{\gamma}} \tau^{\mu}_{a} h_{\gamma \epsilon} h^{\nu \epsilon} h^{\rho \sigma} - 2 \partial_{\mu}{\tau^{\nu}_{a}} \partial_{\rho \sigma}{\xi^{\gamma}} \tau_{\gamma}^{a} h_{\nu \epsilon} h^{\mu \epsilon} h^{\rho \sigma} \nn \\ && + 2 \partial_{\mu}{\tau_{\nu}^{a}} \partial_{\rho \sigma}{\xi^{\gamma}} \tau_{\gamma}^{b} \tau^{\mu}_{a} \tau^{\nu}_{b} h^{\rho \sigma}  - \partial_{\mu}{h_{\nu \rho}} \partial_{\sigma \gamma}{\xi^{\epsilon}} h_{\epsilon \delta} h^{\mu \delta} h^{\nu \rho} h^{\sigma \gamma} - \partial_{\mu}{h_{\nu \rho}} \partial_{\sigma \gamma}{\xi^{\epsilon}} \tau_{\epsilon}^{a} \tau^{\mu}_{a} h^{\nu \rho} h^{\sigma \gamma} \nn \\ && - 2 \partial_{\mu}{\tau_{\nu}^{a}} \partial_{\rho \sigma}{\xi^{\gamma}} \tau^{\nu}_{a} h_{\gamma \epsilon} h^{\mu \epsilon} h^{\rho \sigma} - 2 \partial_{\mu}{\tau_{\nu}^{a}} \partial_{\rho \sigma}{\xi^{\gamma}} \tau_{\gamma}^{b} \tau^{\mu}_{b} \tau^{\nu}_{a} h^{\rho \sigma} - \partial_{\mu}{h_{\nu \rho}} \partial_{\sigma \gamma}{\xi^{\mu}} h^{\nu \sigma} h^{\rho \gamma} \nn \\ && - 2 \partial_{\mu}{h_{\nu \rho}} \partial_{\sigma \gamma}{\xi^{\nu}} h^{\mu \rho} h^{\sigma \gamma} + 2 \partial_{\mu}{h_{\nu \rho}} \partial_{\sigma \gamma}{\xi^{\nu}} h^{\mu \sigma} h^{\rho \gamma} - 2 \partial_{\mu}{\tau_{\nu}^{a}} \partial_{\rho \sigma}{\xi^{\mu}} \tau^{\rho}_{a} h^{\nu \sigma} + 2 \partial_{\mu}{\tau_{\nu}^{a}} \partial_{\rho \sigma}{\xi^{\nu}} \tau^{\rho}_{a} h^{\mu \sigma} \nn \\ && - 2 \partial_{\mu}{\tau_{\nu}^{a}} \partial_{\rho \sigma}{\xi^{\nu}} \tau^{\mu}_{a} h^{\rho \sigma} + \partial_{\mu}{h_{\nu \rho}} \partial_{\sigma \gamma}{\xi^{\mu}} h^{\nu \rho} h^{\sigma \gamma} + 2 \partial_{\mu}{\tau_{\nu}^{a}} \partial_{\rho \sigma}{\xi^{\mu}} \tau^{\nu}_{a} h^{\rho \sigma} - 2 \partial_{\mu}{h_{\nu \rho}} \partial_{\sigma \gamma}{\xi^{\epsilon}} h_{\epsilon \delta} h^{\mu \sigma} h^{\nu \gamma} h^{\rho \delta} \nn \\ && + 2 \partial_{\mu}{\tau^{\nu}_{a}} \partial_{\rho \sigma}{\xi^{\gamma}} \tau_{\gamma}^{a} h_{\nu \epsilon} h^{\mu \rho} h^{\sigma \epsilon} - 2 \partial_{\mu}{\tau_{\nu}^{a}} \partial_{\rho \sigma}{\xi^{\gamma}} \tau^{\rho}_{a} h_{\gamma \epsilon} h^{\mu \sigma} h^{\nu \epsilon} - 2 \partial_{\mu}{\tau_{\nu}^{a}} \partial_{\rho \sigma}{\xi^{\gamma}} \tau_{\gamma}^{b} \tau^{\nu}_{b} \tau^{\rho}_{a} h^{\mu \sigma} \nn \\ && + \partial_{\mu}{h_{\nu \rho}} \partial_{\sigma \gamma}{\xi^{\epsilon}} h_{\epsilon \delta} h^{\mu \delta} h^{\nu \sigma} h^{\rho \gamma} + 2 \partial_{\mu}{\tau_{\nu}^{a}} \partial_{\rho \sigma}{\xi^{\gamma}} \tau^{\rho}_{a} h_{\gamma \epsilon} h^{\mu \epsilon} h^{\nu \sigma} + \partial_{\mu}{h_{\nu \rho}} \partial_{\sigma \gamma}{\xi^{\epsilon}} \tau_{\epsilon}^{a} \tau^{\mu}_{a} h^{\nu \sigma} h^{\rho \gamma} \nn \\ && + 2 \partial_{\mu}{\tau_{\nu}^{a}} \partial_{\rho \sigma}{\xi^{\gamma}} \tau_{\gamma}^{b} \tau^{\mu}_{b} \tau^{\rho}_{a} h^{\nu \sigma} = 0 \, .
\eea
To obtain the last equality one needs to impose that $h_{\mu \nu} h^{\nu \rho} = \delta_{\mu}^{\rho} - \tau_{\mu}{}^{a} \tau^{\rho}{}_{a}$. The covariant rewriting of the Ricci scalar can be checked in \cite{Enew}.
{}
\end{document}